\documentclass[12pt]{article}
\usepackage{epsfig}

\def\beq{ \begin{equation}}
\def\eeq{\end{equation} }
\def\bea{\begin{eqnarray}}
\def\eea{\end{eqnarray}}
\newcommand{\nn}{\nonumber}

\newcommand{\msl}[1]{{{m}}_{\tilde{L}_{#1}}^2}

\begin{document}

\pagestyle{empty}
\begin{flushright}
{CERN-PH-TH/2002-181}\\
hep-ph/0612292\\
\end{flushright}
\vspace*{5mm}
\begin{center}
{\large {\bf CP and Lepton-Number Violation in GUT Neutrino Models with Abelian Flavour Symmetries}} \\
\vspace*{1cm}
{\bf John~Ellis$^1$, Mario~E.~G{\'o}mez$^2$} and {\bf Smaragda~Lola$^3$} \\
\vspace{0.3cm}
$^1$ Theory Division, CERN, CH-1211 Geneva 23, Switzerland \\
$^2$ Departamento de F\'{\i}sica Aplicada, University of Huelva, 21071 Huelva, Spain \\
$^3$ Department of Physics, University of Patras, GR-26500 Patras, Greece \\
\vspace*{2cm}
{\bf ABSTRACT} \\ 
\end{center}
\vspace*{5mm}
\noindent

We study the possible magnitudes of CP and lepton-number-violating quantities in specific GUT models
of massive neutrinos with different Abelian flavour groups, taking into
account experimental constraints and requiring
successful leptogenesis. We discuss SU(5) and flipped SU(5) models 
that are consistent with the present data on neutrino mixing and upper limits on the violations of charged-lepton flavours and explore their predictions for the CP-violating oscillation and Majorana phases. In particular, we discuss string-derived flipped SU(5)
models with selection rules that modify the GUT structure and provide
additional constraints on the operators, which are able to account for the magnitudes of some of the coefficients that are often set as arbitrary parameters in generic Abelian models.

\vspace*{5cm}
\noindent

\begin{flushleft} CERN-PH-TH/2006-181 \\
\end{flushleft}
\vfill\eject

\setcounter{page}{1}
\pagestyle{plain}

\pagebreak

\section{Introduction}
In recent years, several experiments have provided convincing evidence for
neutrino oscillations \cite{S-Kam, SNO, KamLAND, K2K, MINOS},
together with important information on the neutrino mass differences 
and mixing angles \cite{N-data}. It has now been established that the atmospheric neutrino mixing angle, $\theta_{23}$, is
large, with a central value $\sim \pi/4$, so that $0.31 \leq
\sin^2{\theta_{23}}\leq 0.72$. The solar MSW angle $\theta_{12}$ \cite{MSW}
is also large, but not maximal, with $\sin^2{\theta_{12}}\sim
0.3$. Upper limits from the CHOOZ experiment, in particular, establish that
the third mixing angle, $\theta_{13}$, must be small, with a present upper
limit of $\sin^2{\theta_{13}}\leq 0.1$~\cite{CHOOZ}. Finally, experiments have
established the following ranges for the mass differences: $\Delta
m^2_{atm}\sim 2.5~10^{-3}~eV^2$ and $\Delta m^2_{sol}\sim 8~10^{-5}~eV^2$~\cite{N-data}.

These important results leave several questions open for model builders to
consider:\\
~~- Is the atmospheric neutrino mixing indeed maximal?\\
~~- How large is the solar neutrino mixing, and can it be related to 
mixing among quarks?\\
~~- What is the pattern of masses for the neutrinos?\\
~~- What is the magnitude of $\theta_{13}$?\\
~~- Is there significant CP violation in the neutrino and charged-lepton sectors?\\
In this paper we study, in particular, the answers to the last two questions that are provided by
models that propose different answers to the first three questions.

The existence of neutrino masses and oscillations may have further
consequences. For example, the decays of heavy neutrinos may lead to
leptogenesis, which would impose additional constraints on CP-violating
phases and on the magnitudes of certain neutrino Yukawa couplings~\cite{leptorev}. These
should be large enough to generate a sufficient net lepton number, but not
above the value that would erase any generated asymmetry through
equilibriation effects. Also, mixing in the charged-lepton sector is
enhanced by radiative corrections in supersymmetric models that 
mix the sleptons, contributing via loop diagrams to rare decays and flavour
conversions~\cite{BM,KO}.
The rates for such processes mediated by sparticles may be
large and close to the current experimental bounds in many
supersymmetric models. Moreover, the possibility of CP-violation is
crucial in this analysis~\cite{LEP-CP-NU} and additional CP-violating quantities
may be observable in the charged-lepton sector.
The model parameters that would be responsible
for leptogenesis, flavour and CP violation in charged-lepton processes 
are different from those measurable in neutrino oscillation experiments, and
also motivate this paper.

As we discuss below, the predictions for these additional
parameters vary in different models. However, there are also several
correlations. A combined analysis, using all available processes, could
give significant constraints on models, and even exclude certain
classes of models. The joint study of phases, CP and lepton-flavour violation is a central issue
in this programme, and its better understanding is the central goal of
this work.

The fact that the fermion mass matrices exhibit a hierarchical structure
suggests that they are generated by an underlying flavour symmetry for which
the simplest possibility would be an Abelian family symmetry.
In such a case, one can parametrize the charges of the Standard Model
fields under the symmetry as in Table 1.
The Higgs charges are chosen so that the terms $f_3 f^c_3 H$ (where
$f$ stands for a generic fermion, the subscript $3$ denotes the third
family, and $H$ denotes either $H_{1}$ or $H_{2}$) have zero net charge whereas other
terms have non-zero charges.
Then, in general only the (3,3) element of the associated mass matrix will
be non-zero as long as the U(1) symmetry remains unbroken.  
The remaining matrix elements may be generated when the U(1)
symmetry is spontaneously broken \cite{FN,IR} by fields $\theta, \; \bar{\theta}$ 
that are singlets of the Standard Model gauge group, with U(1) charges that
are in most cases taken to be $\pm 1$, respectively. Here we assume them
to have equal vacuum expectation values (vev's). The suppression factor for
each entry depends on the family charge: the higher the net U(1) charge
of a term $f_i f^c_j H$, the higher the power $n$ in the
non-renormalisable term $f_i f^c_j H \left [ ( \frac{\langle \theta \rangle}{M} \right)^n
$ or $( \frac{\langle \bar{\theta} \rangle}{M})^n $] that has zero charge, where
$M$ is a high mass scale that is assumed to be much larger than $\langle
\theta \rangle = \langle {\bar \theta} \rangle$. The $f_3 f^c_3 H$ couplings are not
suppressed, because they have no net charge.

\begin{table}[!ht]
\begin{center}
\begin{tabular}{|c|cccccccc|}
\hline
 & $Q_{i}$ & $u_{i}^{c}$ & $d_{i}^{c}$ & $L_{i}$ & $e_{i}^{c}$ & 
$\nu_{i}^{c} $ & $H_{2}$ & $H_{1}$ \\ 
\hline
U(1) & $\alpha _{i}$ & $\beta _{i}$ & $\gamma _{i}$ & $c_{i}$ & 
$d_{i}$ & $e_{i}$ & $-\alpha _{3}-\beta _{3}$ & $-\alpha _{3}-\gamma _{3}$ 
\\ \hline 
\end{tabular}
\caption{\it Notation for possible U(1) charges of the various
Standard Model fields, where $i$ is a generation index.}
\end{center}
\end{table}

Whilst such a family symmetry provides a promising origin for a
hierarchical pattern of fermion masses, in order to go further it is
necessary to specify the charges of the quarks, charged leptons and
neutrinos \cite{guid}. In principle, one can fit the observed masses and mixing angles
by choices of flavour charges that are not constrained by additional
symmetries. However, the structure of the Standard Model is suggestive of
an underlying unification relating quark and lepton multiplets, as occurs
in various GUT models. In such a case, it is important to try to
understand the different mass and mixing patterns and other correlations
predicted by different GUTs, and the resulting patterns of CP and lepton-flavour violation in
particular. This is the primary focus of this paper. Since particles in the same multiplet of a GUT group have the same
charge, the predictions and correlations between observables will in general be different for different GUTs and charge
assignments.

A crucial question when studying flavour symmetries is whether these are
indeed {\it Abelian}, as described above, or {\it non-Abelian}. In the
case of an Abelian flavour symmetry, taking predictions for the masses and
mixings {\em only} from the flavour and GUT structure and
assuming that unknown numerical coefficients are generically ${\cal O}(1)$ leads
in the simplest and apparently more natural neutrino models to large mass
hierarchies and unacceptably small solar mixing, which is correlated with
the hierarchies of charged-lepton masses \cite{LR}. Large solar mixing, as required
by the experimental data, has therefore to arise either from the right-handed neutrino
sector (as may arise, for instance, in see-saw models with dominance by a single right-handed
singlet neutrino, and other zero-determinant solutions), or by imposing
more U(1) symmetries and/or introducing more fields. However, this
strategy introduces additional model dependence and loses predictivity. On
the other hand, large atmospheric mixing is naturally predicted in a wide
class of Abelian models, and non-Abelian flavour symmetries typically
predict naturally large angles for both solar and atmospheric 
neutrinos~\cite{nonab,nonab-KR}.

Since there is no charge quantisation in Abelian groups, contrary to the non-Abelian case,
the symmetry cannot determine alone the numerical values of the 
mass-matrix elements and hence the mixing angles and phases. Nevertheless, information on the phases and values of the
coefficients can be inferred phenomenologically, and individual Abelian
models may be used to make interesting qualitative predictions, just as
they have for the mixing angles, based on the powers of the small parameters 
$\langle\theta\rangle/M, \langle{\bar \theta} \rangle/M$ that
are allowed by the symmetries. However, any predictivity requires the
existence of stable solutions where a small change in the parameters does
not change drastically the predictions.  Non-Abelian symmetries are more
predictive in this respect. However, once the symmetry is broken (as is
required to generate large lepton mass hierarchies) this predictivity is
to a certain extent lost. Even in this case, therefore, the phases are
derived by a combination of theoretical and phenomenological
considerations, again in a model-dependent way.

The fact that neutrinos
have masses and mix with each other like quarks also implies that, in
principle, we can expect non-negligible CP-violation in the neutrino
sector. This may manifest itself in several different ways in low-energy
neutrino physics, including the Maki-Nakagawa-Sakata (MNS) oscillation
phase $\delta$ and two possible Majorana phases. Additional phases
arise in extensions of the light-neutrino sector to include either heavy
singlet neutrinos and/or charged leptons. For example, even the minimal
three-generation see-saw model has 6 independent CP-violating phases, and
leptogenesis is independent of the light-neutrino phases, in particular. On the other
hand, some of the phases on which leptogenesis does depend may in
principle be observable in the charged-lepton sector, in the presence of
low-energy supersymmetry. As we discuss below, these observables may take very
different values in classes of models that fit the present neutrino data equally well.

The structure of this paper is as follows. In Section 2 we discuss the neutrino and charged-lepton
observables for which we investigate the predictions of various GUTs with Abelian flavour
symmetries. The specific GUT models and their predictions are discussed in Section~3. 
Our conclusions are set out in Section~4, and appendices present details of the specific GUTs
studied.

\section{Observables in the Lepton Sector}

\subsection{Neutrino Observables}

The mixing matrices $V_{e,D}, U_{R,\nu}$ that yield physical CP and 
flavour violating observables are those making the following diagonalisations:
\bea
V_e^T Y_e Y_e^\dagger V_e^*&=&Diag(y_e^2,y_\mu^2,y_\tau^2),\\
V_D^T Y_\nu Y_\nu^\dagger V_D^*&=&Diag(y_\nu^2,y_\nu^2,y_\nu^2),\\
U_R^T M_{RR} U_R &=&Diag(M_1,M_2,M_3),\\
U_\nu^T m_{eff} U_\nu &=&Diag(m_{\nu_1},m_{\nu_2},m_{\nu_3}),
\eea

where $Y_{e,\nu}$ stand for the Yukawa couplings of neutrinos
and charged leptons respectively, and $M_{RR}$ is the heavy Majorana neutrino 
mass matrix, which we assume to be three-dimensional  (for a review 
see \cite{neutrino_rev}). 
The effective neutrino mass matrix is:
\beq
m_{eff} \approx m_D^\nu{1\over M_{RR}} {m_D^\nu}^T .
\eeq

In terms of the above matrices, the Maki-Nakagawa-Sakata (MNS) matrix becomes 
\beq
U_{MNS} \equiv U = V_e U_\nu^\dagger,
\eeq
and can be parametrized as:
\begin{equation}
U= diag(e^{i\delta_e},e^{i\delta_\mu},e^{i\delta_\tau}).V. 
 diag(e^{-i\phi_1/2},e^{-i\phi_2/2},1) ~,
\end{equation}
where
\begin{equation}
V=\left( 
\begin{array}{ccc}
c_{12}c_{13} & s_{12}c_{13} & s_{13}e^{-i\delta}\\
-c_{23}s_{12}-s_{23}s_{13}c_{12}e^{i\delta} &
c_{23}c_{12}-s_{23}s_{13}s_{12}e^{i\delta} &
s_{23}c_{13}\\
s_{23}s_{12}-c_{23}s_{13}c_{12}e^{i\delta} &
-s_{23}c_{12}-c_{23}s_{13}s_{12}e^{i\delta} &
c_{23}c_{13}
\end{array}
\right) ~,
\end{equation}
and $c_{ij}$ and  $s_{ij}$ stand for $\cos\theta_{ij}$ and 
 $\sin\theta_{ij}$, respectively.
In this formalism, the three neutrino mixing angles and the six CP-violating
phases are given by
\begin{flushleft} \begin{eqnarray}
&& \theta_{13}={\rm arcsin}(|U_{13}|), 
~~\theta_{12}={\rm arctan}\left(\frac{|U_{12}|}{|U_{11}|}\right),
~~\theta_{23}={\rm arctan}\left(\frac{|U_{23}|}{|U_{33}|}\right) ~, \nonumber \\
&&\delta_{\mu}={\rm arg}(U_{23}), 
~~\delta_{\tau}={\rm arg}(U_{33}),
~~\delta=-{\rm arg}\left(
\frac{\frac{U_{ii}^*U_{ij}U_{ji}U_{jj}^* }{c_{12}c_{13}^2
c_{23}s_{13}}+c_{12}c_{23}s_{13}}{s_{12}s_{23}}\right) ~, \nonumber \\
&&\delta_e={\rm arg}(e^{i \delta} U_{13}),
~~\phi_1=2{\rm arg}(e^{i \delta_e} U_{11}^*),
~~\phi_2=2{\rm arg}(e^{i \delta_e} U_{12}^*) ~. \nonumber  
\end{eqnarray}
\end{flushleft}
where  $i,j = 1,2,3$ and $i \neq j$. A useful measure of the amount of 
CP-violation in the oscillations of light neutrinos is the Jarlskog invariant
\begin{eqnarray}
J_{CP}&=& \frac{1}{2}|Im(U_{11}^*U_{12}U_{21}U_{22}^*)|= 
\frac{1}{2}|Im(U_{11}^*U_{13}U_{31}U_{33}^*)| \nonumber \\
&=&\frac{1}{2}|Im(U_{22}^*U_{23}U_{32}U_{33}^*)|=
\frac{1}{2}|c_{12}c_{13}^2c_{23}\sin\delta s_{12} s_{13} s_{23}| ~.
\label{Jarlskog}
\end{eqnarray}

It is clear that different models that reproduce the experimental values of 
$\theta_{12,23}$ and the differences in light neutrino masses-squared may well predict different
amounts of CP-violation in neutrino oscillations (\ref{Jarlskog}) and elsewhere. 
So far, experiment has provided only an upper bound on the 
possible magnitude of $\theta_{13}$, and no information on $\delta$. The magnitude of CP-violation in neutrino oscillations may even be zero, for example in models
with texture zeroes in the (1,3) entries. As we see later, specific models
may connect $\theta_{13}$ and $\delta$, and also provide connections to other
observables. We aim here at the differentiation of possible flavour models according to
their predictions for observables in the neutrino sector - what are the expected magnitudes of $\theta_{13}, \delta$ and $\phi_{2,3}$? - and elsewhere~\footnote{In order to calculate mixing angles and CP-violating phases in an
automated way, we analyze the different models using the package REAP
\cite{REAP}.}. We now discuss some of the other observables in more detail.

\subsection{Leptogenesis}

The textures of neutrino Yukawa couplings may be constrained by the requirement of
successful leptogenesis \cite{leptogenesis}, which requires obtaining the correct
magnitude of lepton-flavour violation (LFV) in the 
decays of heavy, right-handed Majorana neutrinos via a difference between the branching ratios (BR) for heavy-neutrino decays into leptons and antileptons:
\begin{eqnarray}
BR(N_L^c \rightarrow  \overline{\Phi}+\ell) \; \neq \; 
BR(N_L^c \rightarrow \Phi+\overline{\ell}). \nonumber
\end{eqnarray}
Since non-perturbative lepton- and baryon-number-violating interactions 
mediated by sphalerons are in thermal equilibrium up to
the time of the electroweak phase transition, a non-zero lepton
number gives rise to a non-zero baryon number, by sharing the lepton
asymmetry $\Delta L \neq 0$ with a baryon asymmetry $\Delta B \neq 0$.

In order to avoid washout of the initial decay asymmetry
by perturbative decay, inverse decay and scattering interactions, the model
must satisfy an out-of-equilibrium condition,
namely that the heavy-neutrino decay rate is smaller than 
the Hubble parameter $H$ at 
temperatures of the order of the right handed neutrino masses. 
The tree-level width of the heavy neutrino $N_i$ with  
mass $M_i$ is:
$\Gamma = [(\lambda^\dagger \lambda)_{ii}/{8 \pi}] M_{i}$, which should 
be compared with the Hubble expansion rate
{ $H \approx 1.7 ~g_*^{1/2} ~{T^2}/{M_P}$
}, (where $g_*^{MSSM} \approx 228.75$, $g_*^{SM} = 106.75$ 
and $M_P$ is the Planck mass),
leading to the requirement
\bea
\frac{(\lambda^\dagger \lambda)_{ii}}{14 \pi g_*^{1/2}} M_{P} < M_{i}.
\nonumber
\eea
Here, $\lambda \equiv Y_\nu^\prime$ denotes the
neutrino Yukawa matrix in the basis where
the Majorana masses $M_{i}$ are diagonal.
This condition may be implemented more accurately by
looking in detail at the Boltzmann equations, but this
formula is sufficient for our purposes.

The CP-violating decay asymmetry $\epsilon_j$ in the decay of a
heavy-neutrino flavour $j$ arises from the
interference between tree-level and one-loop amplitudes, and is
\bea
     \epsilon_j & =  & 
{1\over( 8 \pi Y_{\nu}^{\prime\dagger}Y_{\nu}^\prime)_{jj}}
     \sum_{i\neq j} {\rm Im} 
\left [
((Y_{\nu}^{\prime\dagger} Y_{\nu}^\prime)_{ji})^2
\right ] f \left 
(\frac{M^2_{j}}{M^2_{i}} \right ), 
\label{eq:eps}
\eea
where the kinematic function
\bea
 f(y) & = & \sqrt{y}\left[{1\over 1-y} + 1-(1+y)\ln\left({1+y\over y}
     \right)\right].\nonumber 
    \eea
The first term in $f(y)$ arises from self-energy corrections, while the second
and third terms arise from the one-loop vertex. 
For models with degenerate Majorana masses, 
one expects a resonant enhancement of the lepton asymmetry,
since in this case  $y \sim 1$ and
\begin{equation}
f(y) \sim -{{|M_{i}|} \over{2(|M_{j}| - |M_{i}|)}}.
\label{degenerate}
\end{equation}
This will be manifest in some of the examples below.

The lepton asymmetry produced by the 
decays of the heavy neutrinos is in general diluted partially by 
lepton-number-violating processes~\cite{leptogenesis}.
The washout factor $\kappa_i$ is approximately
\begin{equation}
	\kappa_i(\tilde{m}_i) \simeq 0.3\left({{10^{-3}{\rm eV}}\over
		{\tilde{m}_i}}\right)\left(\log {{\tilde{m}_i}\over
		{10^{-3}{\rm eV}}}\right)^{-0.6} ,
\label{eq:kappa}
\end{equation}
where
\begin{equation}
	\tilde{m}_i = (M'^\dagger_{RR} M'_{RR})_{ii}/|M_i| ,
\label{eq:mtilde}
\end{equation}
and the primes denote the basis where the right-handed neutrino mass
matrix is diagonal.

Then, for a supersymmetric model, the lepton and baryon 
asymmetries are given by 
\begin{eqnarray}
	\left(\frac{n_L}{s}\right) &\simeq& 0.8 \times 10^{-2} 
        \kappa_i \epsilon_i ~\frac{1}{\Delta} ,
		\\
	\left(\frac{n_B}{s}\right) &\simeq& -2.8 \times 10^{-3} 
        \kappa_i \epsilon_i ~\frac{1}{\Delta} ,
		\\
	\eta_B &\simeq& -0.02 \kappa_i \epsilon_i ~\frac{1}{\Delta} ,
\label{eq:susy1}
\end{eqnarray}
where the entropy of the co-moving volume $s$ is given by
$s = (2/45)g_* \pi^2 T^3$, and
$g_* = 228.75$ refers to 
the effective number of relativistic degrees of freedom contributing to the 
entropy in our supersymmetric model.

In the above relations, $\Delta$ is a dilution factor that 
appears due to entropy production from symmetry breaking and
an inflationary phase.
This effect has been studied in~\cite{cehno1},  in the context of
the breaking  of $SU(5)\times U(1)$ when a singlet
field $\Phi$  gets a vev. In this case, the dilution
factor is \cite{JD} 
$\Delta = \frac {s(R_{d\Phi})}{s(R_{d\eta})}
(\frac {R_{d\Phi}}{R_{d\eta}})^3  \sim
\frac {V^3 m_\eta^{3/2}}{{\alpha_\Phi}^{1/2}
\ m_{SUSY}^{3/2} M_P^3}$.
Here, the $\Phi$ decay rate is given by
$\Gamma_\Phi = \alpha_\Phi 
\frac { m^3_{SUSY}}{V^2}$, 
$V$ is the scale where
the vev of the Higgs ${\tt 10}$ and ${\tt \overline{10}}$ 
break the flipped-SU(5) group, $m_\eta$ is the inflaton mass,
and $m_{SUSY}$ is the supersymmetry breaking scale.

Successful leptogenesis would require~\cite{Spergel:2003cb}:
\beq
\eta_B=(6.15\pm 0.25)\times 10^{-10} .
\label{eq:etarange}
\eeq
Clearly, $\kappa_i/\Delta < 1$ in general and a model with too small a value of 
$\epsilon_i$ would not be valid. However, as we discuss below, this is not the
case in the models we study, which generally require $\Delta \gg 1$ in order
to provide a successful scheme of leptogenesis. An alternative to invoking a
large dilution factor would be to adopt a wider range
of neutrino Yukawa couplings and Majorana masses that
can lead to consistent  solutions.

The impact of flavour effects in leptogenesis has also been considered in
\cite{fl-lep1} and more recently in \cite{fl-lep2}, and may affect the results
by factors of $\sim 2$. Since the conclusions of our paper are not affected 
by the possible presence of such factors, we will not proceed with more 
detailed considerations in this direction.

\subsection{Charged-Lepton-Flavour Violation}

In the context of low-energy supersymmetry, 
mixing in the neutrino sector also generates mixing in the sleptons via
loop corrections, contributing to rare decays and lepton-flavour conversions.
We evaluate these effects in the context of the CMSSM, where the soft 
supersymmetry-breaking masses of the charged and neutral sleptons are 
assumed to be universal at the GUT scale, and may be written in diagonal
form with a common value $m_0$. 
Off-diagonal entries in the slepton mass matrix $\msl{}$ are then generated 
radiatively by the renormalization-group evolution from 
the GUT scale $M_{GUT}$.

The branching ratios for LFV decays can be described well by a 
single-mass-insertion approximation~\cite{HIM,LAM}:
\beq
\label{brdef}
BR(l_i\to l_j\gamma)\approx{\alpha^3\over G_F^2} 
\mathcal{F}(m_0,M_{1/2},\mu)|\msl{ij}|^2 \tan^2\beta,
\label{apLFV}
\eeq
where $\mathcal{F}$ is a function of the soft supersymmetry-breaking 
masses fixed at the high scale $M_{GUT}$, and $i\neq j$ are 
generation indices. 
In the commonly-used approximation
for solving the renormalisation-group equations of a single 
intermediate right-handed neutrino threshold, 
the mass corrections  are related to the neutrino 
Yukawa couplings by \cite{EHRS}
\beq
\label{RGEsol1}
\msl{ij} = \kappa \sum_k \bar{Y_\nu}^{ik} (\Delta t+\Delta\ell_k) 
(\bar{Y_\nu}^{jk})^*,
\label{approx}
\eeq
where $\bar{Y}_\nu$ is the Dirac neutrino Yukawa matrix in a basis 
where both the heavy-Majorana and charged-lepton couplings become diagonal,
and
\beq
\kappa =-6m_0^2-2A_0^2, ~~\Delta t=\ln(M_{GUT}/M_3)/16\pi^2, ~~
\Delta\ell_k=\ln(M_3/M_k)/16\pi^2 .
\eeq
where $M_i$ with $i=1,2,3$ denote the Majorana neutrino masses 
assuming $M_1< M_2 < M_3$. 

This result is not 
very accurate, but it is a useful approximation for
obtaining analytical expressions for 
lepton-flavour-violating decay rates~\cite{LFVap}.
In the analysis that follows we assume for simplicity that $A_0=0$. If $A_0 \ne 0$, 
this parameter would also be similarly renormalized, via
\begin{equation}
\left(\delta A_e\right)_{ij} \approx 
-\frac{1}{8\pi^2} A_0 Y_{e_i}({ \bar{Y}_\nu}{ \bar{Y}_\nu^\dagger})_{ij}
\log\frac{M_{GUT}}{M_{i}}. 
\label{renA}
\end{equation}
The extra suppression due to $Y_e$ and
the form of $\kappa$ indicate that, unless $A_0$ becomes much larger than
$m_0$, our conclusions will be qualitatively unchanged.

Another lepton-number-violating observable is
$\mu \rightarrow e$ conversion on a nucleus, which has a rate
\begin{eqnarray}
 R({\mu^+ Ti\rightarrow e^+ Ti})&\approx&
\frac{\alpha}{3\pi}\frac{E_ep_e}{m_{\mu}^2}
\frac{Z F_c^2}{C  f(A,Z)}BR(\mu\rightarrow e\gamma)\nonumber\\
  &\approx & 5.6 \times 10^{-3} BR(\mu\rightarrow e\gamma).
\label{convratio}
\end{eqnarray}
This process is very interesting, despite the relative suppression by about two orders of magnitude,
because of the accuracy possible in future measurements of this process.
Within the framework discussed here, a similar suppression is expected for the decay $\mu \rightarrow 3e$,
\begin{equation}
{\Gamma ( \mu^+ \rightarrow e^+e^+e^-) \over \Gamma ( \mu^+ \rightarrow e^+ \gamma)}
\approx 6 \times 10^{-3}.
\label{approxratio}
\end{equation}
However, the present experimental bound on this branching ratio is relatively weak, and the prospects for significant improvement are more distant. It should be noted, though, that this decay does offer the possibility of observing a T-violating asymmetry, which might be observable if $\mu \to e \gamma$ occurs with a rate close to the present experimental upper limit.
The flavour-violating decays $\tau \to e \gamma$ and $\mu \gamma$ are also potentially interesting.
They are governed by formulae similar to those above, and later we present some results for them.

\section{GUTs with Abelian Flavour Symmetries}

As mentioned in the Introduction, we focus in this work on Abelian flavour symmetries, since they are simple and arise naturally in a wide class of models. Moreover, understanding Abelian symmetries is a first step towards the understanding of non-Abelian symmetries. Indeed, by combining two or more Abelian symmetries and introducing more than one field whose vev determines the expansion parameter of the mass matrices, one can simulate to some extent the picture that one would obtain from a non-Abelian structure. 

\subsection{SU(5)}

The first GUT group we analyze is SU(5), whose minimal matter field content is three families of
$(Q,u^{c},e^{c})_{i} \in {\tt 10}$ representations of SU(5), three families of
$(L,d^{c})_i  \in { \tt \overline{5}}$ representations, and heavy right-handed neutrinos $N_L^c \equiv \nu_R$ in singlet representations. Only two heavy $N_L^c \equiv \nu_R$ fields are needed to provide two non-zero masses for light neutrinos, as required by experiment, and more than three are present in some models. However, in what follows, we assume that there are also just three heavy $N_L^c \equiv \nu_R$ fields. This model has the following properties that are important for our analysis: 

(i) the 
up-quark mass matrix is symmetric, \\
(ii) the charged-lepton mass matrix is the transpose of the 
down-quark mass matrix, which relates the mixing of the left-handed leptons to that of the
right-handed down-type quarks. 
Since the CKM mixing in the quark sector is due to a mismatch between the mixing of the left-handed up- and down-type quarks, it is independent of mixing in the lepton sector.
In particular, in SU(5) the large mixing angle  that is observed in atmospheric
neutrino oscillations can easily be consistent with the observed small $V_{CKM}$ mixing.
Most importantly,\\
(iii) it is clear from the 
SU(5) representation structure that the Abelian flavour charges of 
the fermions $(Q,u^{c},e^{c})_{i}$ in the same ${\tt 10}$ representation must be identical, as must the charges of the $(L,d^{c})_i$ in the same ${ \tt \overline{5}}$ representation. This is the type of correlation that we seek to test by examining model predictions for flavour and CP violation in the 
generalised lepton sector.

There are several possibilities for the Abelian flavour charges, that are motivated by theoretical considerations such as anomaly cancellation \cite{K-SU5} as well as phenomenological 
arguments. Within this framework, one may address questions such as: \\
~~- How close to maximal is the atmospheric mixing? \\
~~- How large is the solar mixing? \\
~~- How large are $\theta_{13}$ and $\delta$? \\
~~- At which level is mixing controlled by the hierarchies of charged-lepton masses,
and how strong is the influence from the heavy right-handed neutrino sector?\\
~~- What statements can be made about other observables?

There are several viable sets of textures that we analyze for answers to these questions. 
Taking into account the above-mentioned constraints arising from the SU(5) multiplet structure
(symmetric up-quark mass matrix, and $M_\ell$ the transpose of $M_{down}$), 
the mass matrices are constrained to the forms
\begin{equation}
{\cal M}_{u}\propto 
\left( 
\begin{array}{ccc}
{\epsilon }^{|2x|} & {\epsilon }^{|x+b|} & {\epsilon }^{|x|} \\ 
{\epsilon }^{|x+b|} & \epsilon ^{|2b|} & \epsilon ^{|b|} \\ 
\epsilon ^{|x|} & \epsilon ^{|b|} & 1
\end{array}
\right),
 ~~M_{down}\propto \left( 
\begin{array}{ccc}
{\bar{\epsilon}}^{|x+y|} & {\bar{\epsilon}}^{|x+a|} & {\bar{\epsilon}}^{|x|}
\\ 
{\bar{\epsilon}}^{|y+b|} & \bar{\epsilon}^{|a+b|} & \bar{\epsilon}^{|b|} \\ 
\bar{\epsilon}^{|y|} & \bar{\epsilon}^{|a|} & 1
\end{array}
\right), 
~~M_{\ell }\propto 
\left( 
\begin{array}{ccc}
{\tilde{\epsilon}}^{|x+y|} & {\tilde{\epsilon}}^{|y+b|} & {\tilde{\epsilon}}%
^{|y|} \\ 
{\tilde{\epsilon}}^{|x+a|} & \tilde{\epsilon}^{|a+b|} & \tilde{\epsilon}%
^{|a|} \\ 
\tilde{\epsilon}^{|x|} & \tilde{\epsilon}^{|b|} & 1
\end{array}
\right) ,
\end{equation}
where $a=Q_{2}^{\overline{5}}-Q_{3}^{\overline{5}}$,
$b=Q_{2}^{10}-Q_{3}^{10}$, $x=Q_{1}^{10}-Q_{3}^{10}$, 
 $y=Q_{1}^{\overline{5%
}}-Q_{3}^{\overline{5}}.$
We note that, by assumption, the (3, 3) entries in all the mass matrices are ${\cal O}(1)$,
as would be suitable for large $\tan\beta$~\footnote{We look below also at examples with small 
$\tan\beta$.}, and we have assumed that 
the flavour charges of the two supersymmetric Higgs fields are the same. Since 
we know that the (3,3) entry in the charged-lepton mass matrix is the largest one, we have simplified our considerations by taking a zero Higgs flavour charge~\footnote{The conditions for anomaly cancellation give further insight into the possible textures, but detailed model building goes beyond the scope of this paper.}.

\subsection{An example with large $\tan\beta$}

An interesting possibility is that
{\it maximal atmospheric mixing arises from the charged-lepton sector}.
In this case, the (2,3) and (3,3) lepton entries are comparable, and
$a=0$. The resulting mass matrices have been discussed in the literature,
e.g., in~\cite{LR}. The choice $|b|=2$ leads to the correct
$m_s/m_b$ ratio, while requiring correct (1,2) quark mixing fixes $|x| = 3$
in the down mass matrix~\footnote{We cannot obtain the (1,2) quark mixing from the up sector,
as this would lead to an unacceptably large mass for the up quark.} and
$|y|$ is fixed by the down and charged-lepton mass hierarchies. Thus, one obtains finally
mass matrices of the form \cite{LR,SU5-a}:
\begin{equation}
{\cal M}_{u}\propto  \left( 
\begin{array}{ccc}
\bar{\epsilon}^{6} & \bar{\epsilon}^{5} & \bar{\epsilon}^{3} \\ 
\bar{\epsilon}^{5} & \bar{\epsilon}^{4} & \bar{\epsilon}^{2} \\ 
\bar{\epsilon}^{3} & \bar{\epsilon}^{2} & 1
\end{array}
\right) ,
~~M_{down}\propto 
\left( 
\begin{array}{ccc}
\bar{\epsilon}^{4} & \bar{\epsilon}^{3} & \bar{\epsilon}^{3} \\ 
\bar{\epsilon}^{3} & \bar{\epsilon}^{2} & \bar{\epsilon}^{2} \\ 
\bar{\epsilon} & 1 & 1
\end{array}
\right) ,
~~M_{\ell }\propto  \left( 
\begin{array}{ccc}
\bar{\epsilon}^{4} & \bar{\epsilon}^{3} & \bar{\epsilon} \\ 
\bar{\epsilon}^{3} & \bar{\epsilon}^{2} & 1 \\ 
\bar{\epsilon}^{3} & \bar{\epsilon}^{2} & 1
\end{array}
\right) ,
\label{EX1}
\end{equation}
where a single expansion parameter has been 
used to reproduce the fermion mass hierarchies~\footnote{This example gives an up-quark mass
that tends to be somewhat too high, but this defect can be remedied in specific models.}.

As is clear from the form of the charged-lepton mass matrix, 
unlike what happens for the atmospheric mixing, the solar
mixing cannot be obtained from the charged-lepton sector.
In this case, it has to arise from the neutrino sector, exploiting the see-saw structure.
It is interesting to observe the {\it a priori} large 
value of $\theta_{13}$ given by the charged-lepton sector - indeed, $\theta_{13}$ may 
even be too large - but
the overall (1,3) lepton mixing can be reduced 
either by a numerical coefficient, or by a cancellation between the 
charged-lepton and neutrino mass matrices.
In fact,  comparing the  textures for maximal and non-maximal mixing,
we see that, due to the different magnitudes of the neutrino
mixing angles entering in the Jarlskog invariant, one can expect in principle
different inter-correlations and predictions for CP-violation,
even before looking in 
detail at the heavy right-handed neutrino sector and its
influence on $m_{eff}$.

In general, of course, both the mass structure and the mixings
of neutrinos are more
complicated, because of the heavy Majorana masses of the
right-handed components. We assume that these arise
from terms of the form
$\nu_R\nu_R\Sigma$,  where $\Sigma$ is an SU(3)$\times$
SU(2)$\times$U(1)-invariant Higgs field with $I_W=0$ and a non-zero flavour U(1) 
charge~\footnote{Note that $\Sigma$ is a singlet that enters only in the heavy Majorana mass 
textures. It is not the same as the $\theta$ field that generates the light fermion masses.}.
The possible choices for the
$\Sigma$  charge give a discrete spectrum of forms for the Majorana mass matrix
$M_{\nu_R}$. In the case of a single $\Sigma$ field and assuming a zero Higgs charge and
left-handed lepton charges as above, we have: 
\begin{eqnarray}
m^\nu_D & \propto &  \left[
  \begin{array}{ccc}
    \epsilon^{|y+ n_1|} & \epsilon^{|y+ n_2|} & \epsilon^{|y+ n_3|} \\
    \epsilon^{|a+n_1|} & \epsilon^{|a+n_2|} & \epsilon^{|a+n_3|} \\
    \epsilon^{|n_1|} & \epsilon^{|n_2|} & \epsilon^{|n_3|} 
  \end{array}
\right] , \\
M_{RR}& \propto & \left[
\begin{array}{ccc}
\bar{\epsilon}^{|2n_1+\sigma|}&\bar{\epsilon}^{|n_1+n_2+\sigma|}&\bar{\epsilon}^{|n_1+n_3+\sigma|}\\
\bar{\epsilon}^{|n_1+n_2+\sigma|}&\bar{\epsilon}^{|2n_2+\sigma|}&\bar{\epsilon}^{|n_2+n_3+\sigma|}\\
\bar{\epsilon}^{|n_1+n_3+\sigma|}&\bar{\epsilon}^{|n_2+n_3+\sigma|}&\bar{\epsilon}^{|2n_3+\sigma|}\\
\end{array}
\right],
\end{eqnarray}
where the charges $n_i$ are the U(1) charges of the right-handed neutrinos, and
$\sigma$ is the U(1) charge  of the field $\Sigma$. Unlike what happens for the charged leptons, where the (3,3) entry is the largest one,
the large entry in  $M_{RR}$ can be in any position in the matrix, 
depending on the relative charges of $\Sigma$ and the right-handed neutrinos.

We recall that the effective light-neutrino mass matrix is given by
\[
m_{eff}\approx m_D^\nu{1\over M_{RR}} {m_D^\nu}^T .
\]
Its diagonalization relative to the charged-lepton mass matrix
(which is the transpose of the down-type quark mass matrix) 
determines the MNS mixing matrix.
In the case of a zero flavour charge for the Higgs, the mass 
matrix is determined by the flavour
charges of the left-handed neutrinos, which in SU(5) are the same as those of the
right-handed down quarks.
The effective light-neutrino mass matrix then
has the structure
\beq
m_{eff }\propto  \left( 
\begin{array}{ccc}
{\bar{\epsilon}}^{|2y|} & {\bar{\epsilon}}^{|y+a|} & {\bar{\epsilon}}^{|y|}
\\ 
{\bar{\epsilon}}^{|y+a|} & \bar{\epsilon}^{|2a|} & \bar{\epsilon}^{|a|} \\ 
\bar{\epsilon}^{|y|} & \bar{\epsilon}^{|a|} & 1
\end{array}
\right) .
\end{equation}
From the values of $a$ and $y$ obtained from the charged-fermion mass hierarchies, we then
conclude that
\beq
m_{eff }\propto  \left( 
\begin{array}{ccc}
{\bar{\epsilon}^2} & {\bar{\epsilon}} & {\bar{\epsilon}}
\\ 
{\bar{\epsilon}} & 1 & 1 \\
\bar{\epsilon} & 1 & 1
\end{array}
\right),
~m^\nu_D \propto \left(
  \begin{array}{ccc}
    \bar\epsilon^{|1 \pm n_1|} & \bar\epsilon^{|1 \pm n_2|} & \bar\epsilon^{|1 \pm  n_3|} \\
    \bar\epsilon^{|n_1|} & \bar\epsilon^{|n_2|} & \bar\epsilon^{|n_3|} \\
    \bar\epsilon^{|n_1|} & \bar\epsilon^{|n_2|} & \bar\epsilon^{|n_3|} 
  \end{array}
\right) ,
\end{equation}
which can potentially lead to large solar and atmospheric
neutrino mixing (the expansion parameter of the Dirac neutrino
mass matrix is in principle similar to that for the up quarks, 
since these particles couple to the same Higgs field, but there
can be deviations from this).

A working example along these lines can be 
obtained by the following
choice of charges:
\beq
n_1=2, n_2=-1, n_3=1, \sigma=-1 .
\eeq
Using the indicative choices of the coefficients  
$a^x_{ij}$ shown in Table~\ref{SU5-ltb}, we have calculated the relevant observables 
shown in Table~\ref{largetbvalues}.
The expansions in $\bar{\epsilon}$ are obtained by using similar 
techniques as in Ref.~\cite{King:2002nf}. 
Altering the coefficients $a^x_{ij}$ would affect the numerical values appearing 
in Table~\ref{largetbvalues}, but
not the powers of $\bar{\epsilon}$.


\begin{table}[!ht]
\begin{center}
\begin{tabular}{|l||l|}
\hline
& Parameters in an SU(5) model with large $\tan\beta$   \\ \hline \hline
Charged leptons &
$a^e_{12}= 0.6, a^e_{13}= 0.9,
a^e_{22}= 1.2,  a^e_{23} = -0.5e^{i \pi/3}, 
 a^e_{31}=0.7, a^e_{32}=0.6, a^e_{33} = 0.4$ \\
\hline
$m^\nu_D$ &
$a^\nu_{12} =1.3, a^\nu_{21} =-1.3,  
 a^\nu_{22}= 0.7,
a^\nu_{23} = 1.8 e^{i \pi/5}, 
a^\nu_{32}= 0.7, a^\nu_{33}= 0.5 $ \\
\hline
$M_{RR}$ &
$a^N_{22} = 1, a^N_{33} =1.8$ \\
\hline 
\end{tabular}
\caption{\it Choice of coefficients that reproduce the fermion data for an SU(5) model 
with large $tan\beta$. Coefficients not listed in the Table are set to unity.}
\label{SU5-ltb}

\vspace*{0.8 cm}

\begin{tabular}{|l|l|l|}
\hline

Observables &  Series Expansions & Numerical Values \\ \hline \hline
$m_{\nu_3}$ (eV)  & 
$0.23 \bar{\epsilon}+0.35\bar{\epsilon}^{3} +0.66 \bar{\epsilon}^{5}$ & 0.05(0.047)  
\\ \hline 

$m_{\nu_1}/m_{\nu_3}$  & 
$0.85 \bar{\epsilon}-0.51\bar{\epsilon}^{2} -1.92 \bar{\epsilon}^{3}$  & 0.13(0.17)  \\ 
\hline 

$m_{\nu_2}/m_{\nu_3}$ &  
$0.84\bar{\epsilon}+0.52\bar{\epsilon}^2-1.93 \bar{\epsilon}^{3}$ & 0.17(0.17) \\ \hline 
\hline 

$M_{1}/M_3$  & 
$1.8\bar{\epsilon}-1.21\bar{\epsilon}^{2} -0.34\bar{\epsilon}^{3}$ & $ 0.31(0.36)$  \\ \hline

$M_2/M_3$ &  
$1+1.21 \bar{\epsilon}^{2}+3.62\bar{\epsilon}^{3}$ & $0.95(1)$  \\ \hline 
\hline 

$m_{\nu_{ee}}$ (eV) &
$0.56 \bar{\epsilon}^{3}-2.96\bar{\epsilon}^{5}+12.94 \bar{\epsilon}^{7}$ & 
0.0043(0.0044) \\ \hline 
\hline 

$\theta_{23}$  
& $0.92 -1.39 \bar{\epsilon}^{2}-1.65 \bar{\epsilon}^{4}$ & 0.85(0.92) \\ \hline

$\theta_{12}$  
& $0.78  -1.43\bar{\epsilon}+3.62 \bar{\epsilon}^{3}$ & 0.53(0.78) \\ \hline 

$\theta_{13}$ 
& $0.96 \bar{\epsilon} - 2.92 \bar{\epsilon}^{3}+3 \bar{\epsilon}^{5}$ & 
 0.17(0.19) \\ \hline

$\delta$ 
& $3.72- 0.18 \bar{\epsilon}+6.1 \bar{\epsilon}^{2}$ & 4.3(3.72) \\ \hline 
\hline 

$J_{CP}$ 
& $0.093\bar{\epsilon}-0.26\bar{\epsilon}^{3}-0.57\bar{\epsilon}^{5}$ & 0.017(0.019) \\ \hline \hline 

$\phi_1$ 
& $5.6-3.43\bar{\epsilon}+19.3 \bar{\epsilon}^{2}$ & 5.71(5.6) \\ \hline  

$\phi_2$ 
& $2.46+2.68\bar{\epsilon}+19.3 \bar{\epsilon}^{2}$ & 3.93(2.46) \\ \hline 
\hline 

${\epsilon}_1$ 
& $0.086\bar{\epsilon}^{3}-0.04 \bar{\epsilon}^{4}+0.3\bar{\epsilon}^{5}$ 
& $4.1\times 10^{-4}(0.69\times 10^{-4})$ \\ \hline 

${\epsilon}_2$ 
& $-0.005\bar{\epsilon}^{3}+0.02\bar{\epsilon}^{4}
+{0.026\bar{\epsilon}+0.33 \bar{\epsilon}^{2}\over  2.42+7.24 \bar{\epsilon}}$ 
&0.007 (0.0048) \\ \hline 

${\epsilon}_3$ 
& $0.005\bar{\epsilon}^{3}-0.02\bar{\epsilon}^{4}
+{0.026\bar{\epsilon}+0.33 \bar{\epsilon}^{2}\over  2.42+7.24 \bar{\epsilon}}$ 
&0.0063 (0.0048) \\ \hline
 
\multicolumn{1}{|l|}
 {${\eta}_B\times \Delta$ }&
 \multicolumn{1}{|c|}{$ - $ } &
 \multicolumn{1}{|l|}{
 $3.16 \times 10^{-7}(2.46\times 10^{-7})$ }
\\ \hline \hline 

$\frac{16 \pi^2}{\kappa}\times m^2_{\tilde{L}_{12}}$ 
&$13.1 \bar{\epsilon}-1.42 \bar{\epsilon}^{2}-60 \bar{\epsilon}^{3}$ 
&$2.2(2.6)$ \\ \hline 

 \multicolumn{1}{|l|}
 {$BR(\mu\rightarrow e \gamma)\times \frac{1}{(tan\beta)^2}$ 
 }&
 \multicolumn{1}{|c|}{$ - $ } &
 \multicolumn{1}{|l|}{
 $5.3 \times 10^{-11}(7.4\times 10^{-11})$ }
\\ \hline \hline

$\frac{16 \pi^2}{\kappa}\times m^2_{\tilde{L}_{13}}$ 
&$11.5 \bar{\epsilon}+0.84 \bar{\epsilon}^{2}-41 \bar{\epsilon}^{3}$ 
&$2.2(2.3)$ \\ \hline

 \multicolumn{1}{|l|}
 {
$BR(\tau \rightarrow e \gamma)\times \frac{1}{(tan\beta)^2}$ 
 }&
 \multicolumn{1}{|c|}{
 $ - $ } &
 \multicolumn{1}{|l|}{
$4.4\times 10^{-11}(5.7\times 10^{-11})$ 
 }
\\ \hline \hline 

$\frac{16 \pi^2}{\kappa}\times m^2_{\tilde{L}_{23}}$ 
&$12.7\bar{\epsilon}^{2}-2.2 \bar{\epsilon}^{3}- 51\bar{\epsilon}^{4} $ 
&$0.43(0.51)$ \\ \hline

 \multicolumn{1}{|l|}
 {
$BR(\tau\rightarrow \mu \gamma)\times \frac{1}{(tan\beta)^2}$ 
 }&
 \multicolumn{1}{|c|}{$ - $ } &
 \multicolumn{1}{|l|}{
$2\times 10^{-12}(2.8\times 10^{-12})$ }
\\ \hline \hline 

\end{tabular}

\caption{\it Values of observables predicted in the SU(5) model with 
large $\tan\beta$. The rate of convergence of each expansion  in $\bar{\epsilon}$ can be judged
from the relative magnitudes of the expansion coefficients and
by comparing the exact numerical value of the observables with that
obtained by keeping only the dominant term
in each expansion (given in parenthesis
except for $\epsilon_2$ and $\epsilon_3$, where the 
number in parenthesis corresponds to the full fraction).}

\label{largetbvalues}
\end{center}

\end{table}


\begin{figure}[!ht]
\begin{center}
\epsfig{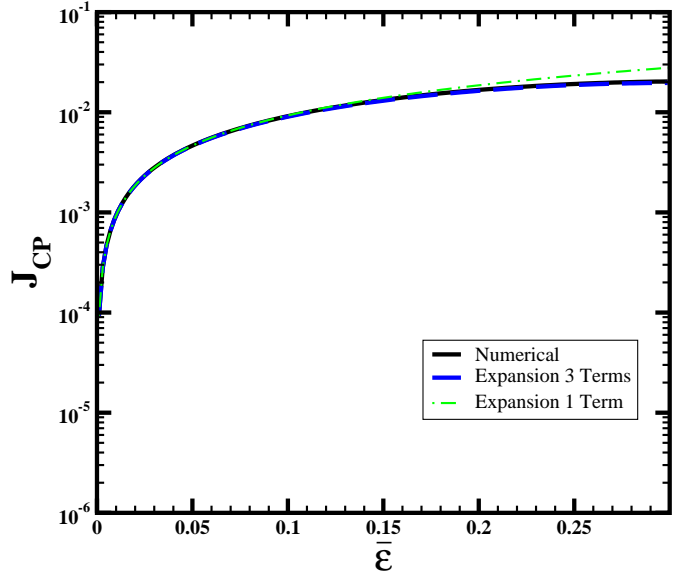}
\end{center}
\vspace*{0.5 cm}
\caption{\it The invariant $J_{CP}$ as a function of the expansion parameter 
$\bar{\epsilon}$ for the case of SU(5) with large $\tan\beta$. The solid 
line was obtained numerically while the green dot-dashed (blue dashed) line uses only  
the first term (three first terms) in the power expansion  in $\bar{\epsilon}$.}
\label{fig:Jarlskog}
\end{figure}


\begin{figure}[!ht]
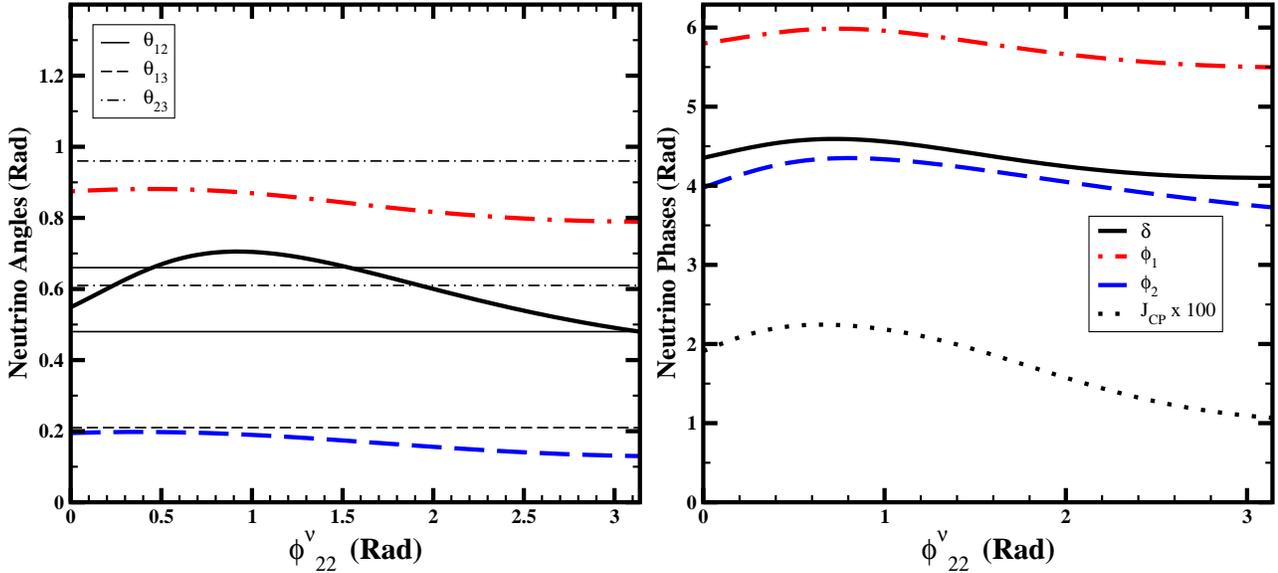

\begin{center}
\epsfig{file=su5-angles.eps,height=3.0in}
\epsfig{file=su5-phases.eps,height=3.0in}
\end{center}
\vspace*{0.5 cm}
\caption{\it The neutrino mixing angles (left) and the phases and 
 values of the CP-violating Jarlskog invariant $J_{CP}$ (right), as functions of the phase $\phi^\nu_{22} \equiv Arg(a_{22}^\nu)$.
The experimental limits on the three neutrino angles
are denoted by horizontal lines.
The values of the table match the values of the figure at  $Arg(a_{22}^\nu)=0$.}\label{fig:largeangles}
\end{figure}

In this example,
$Y_\tau(M_{\rm{GUT}})\sim 0.6 $, which
can be compatible with $\tan\beta$ from 40 to 50
(taking into account the potentially large, model-dependent 
corrections to  $m_b$ and  $m_\tau$ \cite{tanb}).
Moreover, a mass of $m_{\nu_3} \sim 0.05$~eV is compatible with heavy Majorana masses
$\sim 10^{14}$~GeV. Specifically, for $m_{\nu_3}=0.05$~eV, the coefficients in Table~\ref{SU5-ltb}
would predict $M_3 = 2.62 \times 10^{14}$~GeV.
We note the following points of interest.

(i) For almost all entries there is good
agreement between the precise numerical
values and the dominant term in the expansions.
(In the Tables, for  presentation purposes, we replace $\log{\bar{\epsilon}}$ with its 
numerical value at $\bar{\epsilon}=0.2$ in the entries for 
leptogenesis and charged-lepton-flavour violation, which is a good approximation 
for the values of $\epsilon$ of physical relevance). 
As seen in Fig.~\ref{fig:Jarlskog}
(where we compare the numerical values of $J_{CP}$ with those
obtained by keeping the first and the three first terms in the expansions)
the series expansion for $J_{CP}$ is very accurate in the 
relevant range of $\bar{\epsilon}$.

(ii) The model prediction for $\theta_{13}$ is numerically smaller than $\theta_{12}$, but not parametrically smaller. Correspondingly, as seen in Fig.~\ref{fig:largeangles} (left), the model
prediction for $\theta_{13}$ lies not far below the present experimental upper limit. We also
see in this plot that the model predictions for $\theta_{23}$ and $\theta_{12}$ lie
generally near or within the bands allowed by experiment, whatever the value of the model
phase parameter $\phi^\nu_{22}$.

(iii) The model prediction for $\delta$ is parametrically large and ${\cal O}(1)$, as seen in
Fig.~\ref{fig:largeangles} (right). We also see that the light-neutrino Majorana phases $\phi_{1,2}$
are also generically large, and do not vary strongly with $\phi^\nu_{22}$.

(iv) Observable CP-violation is expected in neutrino oscillations, since the Jarlskog invariant
scales as $\bar{\epsilon}$. Numerically, it is typically ${\cal O}(10^{-2})$, as
also seen in Fig.~\ref{fig:largeangles} (right).

(v) As seen in Table~\ref{largetbvalues}, leptogenesis 
is naturally embedded in the model, and is
resonantly enhanced via self-energy corrections,
due to the quasi-degeneracy of the heavy Majorana neutrinos: $M_2 \sim M_3$.
In this example, the self-energy contributions are almost an
order of magnitude larger than the vertex corrections.
The resonant behaviour is also manifest in the expansions, which are
divergent when keeping only the dominant terms
(due to a term of the form ${1\over 1+b\epsilon}$, 
which diverges for $\epsilon \to - 1/b$). To demonstrate this, 
the expansions are displayed by writing separately
the convergent (vertex) and the divergent (self-energy) contributions. As 
discussed above, in order o achieve successful leptogenesis, 
$\eta_b$ must be in the range 
of eq.(\ref{eq:etarange}). For the values presented 
in Table~\ref{largetbvalues}, this implies a dilution factor 
of  $\Delta\sim 500$. Alternatively, by scaling  $m_D^\nu$ with global 
factor $f_s$, we could 
allow for different values of the dilution factor $\Delta$. 
In this case, the 
prediction  $m_{\nu_3} \sim 0.05$~eV could be preserved by also rescaling
$M_{RR}$ by a factor of $f_s^2$. Then, eq.~(\ref{eq:eps}) 
indicates that $\epsilon$  would be modified by a factor $f_s^2$. 
For the set of textures discussed in this section, by
setting   $f_s=0.0441$ one could obtain $\eta_B$ in the range of 
eq.~(\ref{eq:etarange}) with $\Delta\sim 1$ and 
$M_3 \sim  5 \times 10^{11}$~GeV. 
In this case, the predictions 
for lepton flavor violation would be modified by a factor of 
approximately  $f_s^2$. For instance, the previous value of 
$f_s$ would imply a prediction of  BR($\mu \rightarrow e \gamma)\sim 10^{-15} 
tan\beta^2$ which is still of the experimental interest for the large 
values of $\tan\beta$ assumed in this model, while the predictions for 
BR($\tau \rightarrow \mu \gamma$) and BR($\tau \rightarrow e \gamma$) would
become  very small. 

(vi) As also seen in Table~\ref{largetbvalues}, the model predicts a large rate
for $BR(\mu\rightarrow e \gamma)$, close to the present experimental 
upper limit,
as could have been expected for $\tan\beta \geq 40$. This and the 
other rates for lepton-flavour violation have been 
estimated  using (\ref{apLFV}). The values of the soft 
terms are such that the function 
$\mathcal{F}(m_0,M_{1/2},\mu)$ displays a small 
variation with $m_0$, as seen in Fig.~1 
of \cite{LFVap} for $A_0=0$, $M_{1/2}=600$~GeV, 
$m_{0}=300$~GeV. In this case, 
$\mathcal{F}\sim 10^{-18} G_F^2/\alpha^3$. 
This function could increase by two orders 
of magnitude if $M_{1/2}$ were decreased, but this range of 
values is excluded by cosmology. 
Alternatively, the rate could be decreased by 
either increasing $M_{1/2}$ and $m_0$, or by
reducing the Dirac neutrino Yukawa couplings,
with a corresponding adjustment of the heavy Majorana masses.

(vii) More unexpected is the suppression
of   $BR(\tau \rightarrow \mu \gamma)$, and especially 
of $BR(\tau \rightarrow e \gamma)$. These  suppressions arise from the form of the
charged-lepton  mixing matrix $V_e$ which, 
in combination with the degeneracy of the heavy
neutrinos,  tends to align the rotated Yukawa matrix
$Y^\nu$ in the base in which charged leptons are diagonal:
 $\bar{Y}^\nu=V_e^T Y^\nu U_N$.
We return to this issue in subsequent examples.

\subsection{Examples with Small $\tan\beta$}

In \cite{K-SU5}, the U(1) charges were chosen so as to cancel anomalies,
and several solutions were found in the small-$\tan\beta$ regime of supersymmetric theories.
Restricting their attention to solutions with $c_2 = c_3$ and textures where
the heavy-Majorana mass matrix can be considered as diagonal, the authors
of~\cite{K-SU5} found five different fits to 
SU(5) models, whose detailed formulae are given in Appendix I. 
The choices  of U(1) charges correspond to 
different solutions of the anomaly-cancellation 
conditions, and the behaviours of the fits in~\cite{K-SU5}
are very dependent on this. We distinguish:

{\em (a) Fits 1-3:} 
In these fits, single  right-handed neutrino 
dominance (SRHND) has been imposed on $M_{RR}$. 
This is achieved by choosing the charge 
$n_1$ to be a 
negative number between $-\sigma/2$ and 0. Using the GUT values 
found in Table 11 of~\cite{K-SU5}, 
we obtain the best fits when $n_1$ is close to zero, as seen in
Table~\ref{points}.

{\em (b) Fits 4,5:} 
In these cases, we do not find 
fits with clear-cut SRHND. As 
one can see in Table~\ref{points}, the required value for $n_1$ is still 
zero~\footnote{In principle, one could also investigate what happens for other forms of 
$M_{RR}$, since in general  there might be dominant off-diagonal terms
(depending on the singlet charges). However, such a study would go beyond the scope of this paper.}.

We have calculated the relevant observables, 
including the expected CP and charged-lepton-flavour violation,
for representatives of these two classes of fits,
in order to compare their behaviours 
for different sets of parameters within the same GUT framework. 
Since Fits 1-3 and 4-5 have common characteristics, we focus 
on one solution from each group, and work specifically with Fits 2 and 4 of~\cite{K-SU5}.
The values of $\epsilon, \sigma$ and $n_1$ used in these fits are given in Table~\ref{points},
and the chosen values of the numerical coefficients $a_{ij}$ are tabulated in Tables~\ref{var1} 
and ~\ref{var2}
for Fits~2 and 4 of \cite{K-SU5}, respectively. The corresponding predictions for neutrino masses 
and mixing angles, CP and lepton-flavour violation are given in Tables~\ref{Fit2-obs}
and \ref{Fit4-obs}, respectively.

The values of $\tan\beta$ for Fits~2 and 4 are obtained by using the Yukawa couplings
$Y_e$ and the experimental value of $m_\tau$; these lead to
$\tan\beta=2.04$ for Fit~2 and  $\tan\beta=5.15$
for Fit~4. The textures $Y_\nu$ and $M_{RR}$ are scaled so that 
$m_{\nu_3}=0.05$~eV when the largest right-handed neutrino mass becomes 
$M_3=5\cdot 10^{14}$~GeV. For Fit~2, this implies that
\beq 
Y_\nu\rightarrow Y_\nu\cdot 0.46\cdot \epsilon^{-9/2}, \; \; \;  
M_{RR} \rightarrow M_{RR}/\left<\Sigma\right> \cdot 
2.1 \cdot 10^{14} \cdot  \epsilon^{-19/2},
\eeq
while for Fit~4
\beq 
Y_\nu\rightarrow Y_\nu \cdot 0.42, \; \; \;  
M_{RR}\rightarrow M_{RR}/\left<\Sigma\right> \cdot 
2.34 \cdot 10^{14}.
\eeq
The expansions of the observables in powers of $\epsilon$ are obtained as 
discussed above.

\begin{figure}[!ht]
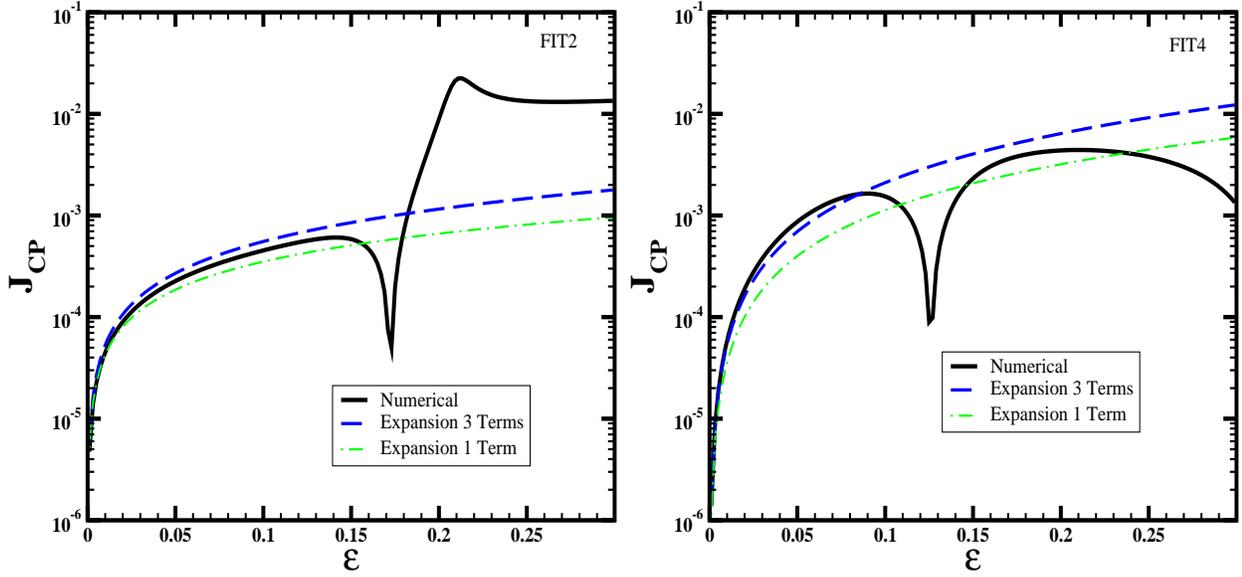

\begin{center}
\epsfig{file=f2-jcp.eps,height=3.0in,width=3.2in}
\epsfig{file=f4-jcp.eps,height=3.0in,width=3.2in}
\end{center}
\vspace*{0.5 cm}
\caption{\it The Jarlskog invariant $J_{CP}$ as a function of the expansion parameter 
$\epsilon$ for two SU(5) models at low $\tan\beta$. The textures correspond to Fits 2 and 4 
of~\cite{K-SU5} for the values of the coefficients $a_{ij}$ in Tables~\ref{var1} and
\ref{var2}. The solid line was obtained numerically while the dot-dash (dash) line uses only  
the first term (three first terms) in the power expansion  in 
$\epsilon$.}
\label{fig:fits-jcp}
\end{figure}

\begin{figure}[!ht]
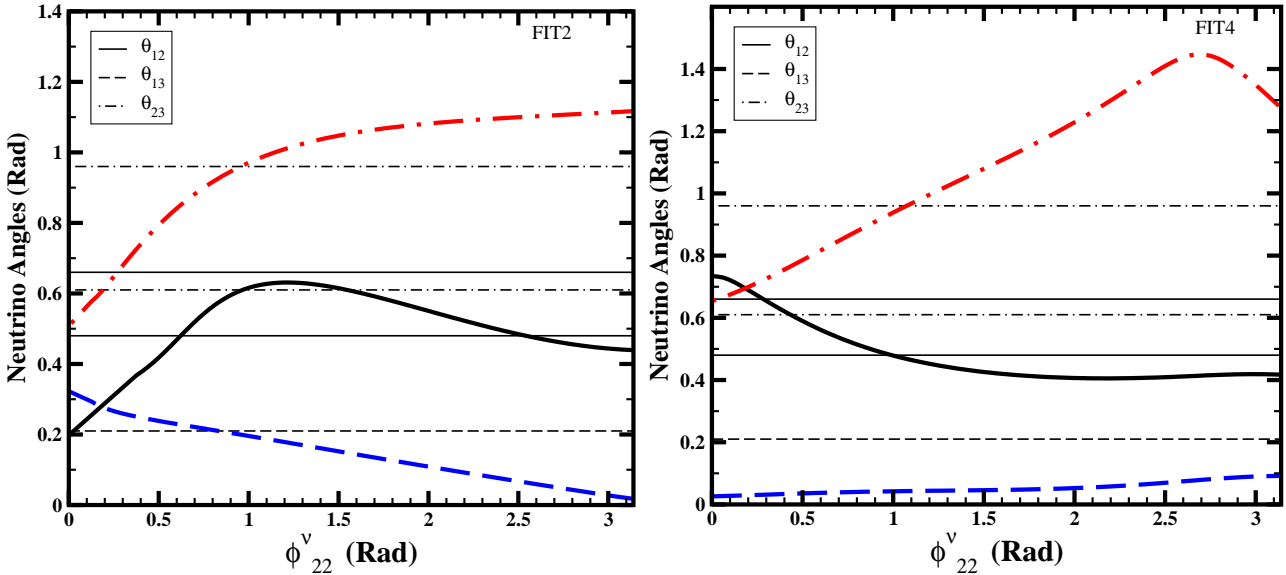

\hspace*{-.35in}
\begin{minipage}{7in}
\epsfig{file=f2-angles.eps,height=3.0in}
\epsfig{file=f4-angles.eps,height=3.0in} 
\hfill
\end{minipage}
\vspace*{0.5 cm}
\caption{\it The neutrino mixing angles $\theta_{23, 12, 13}$ in various SU(5) models, 
as functions of the phase $\phi_{22}^\nu \equiv Arg(a_{22}^\nu)$.
The textures correspond to Fits 2 and 4 of~\cite{K-SU5}
for the values of the coefficients $a_{ij}$ in tables \ref{var1} and
\ref{var2}. The experimental limits on the three neutrino angles
are denoted by horizontal lines.}
\label{fig:fits}
\end{figure}

\begin{figure}[!ht]
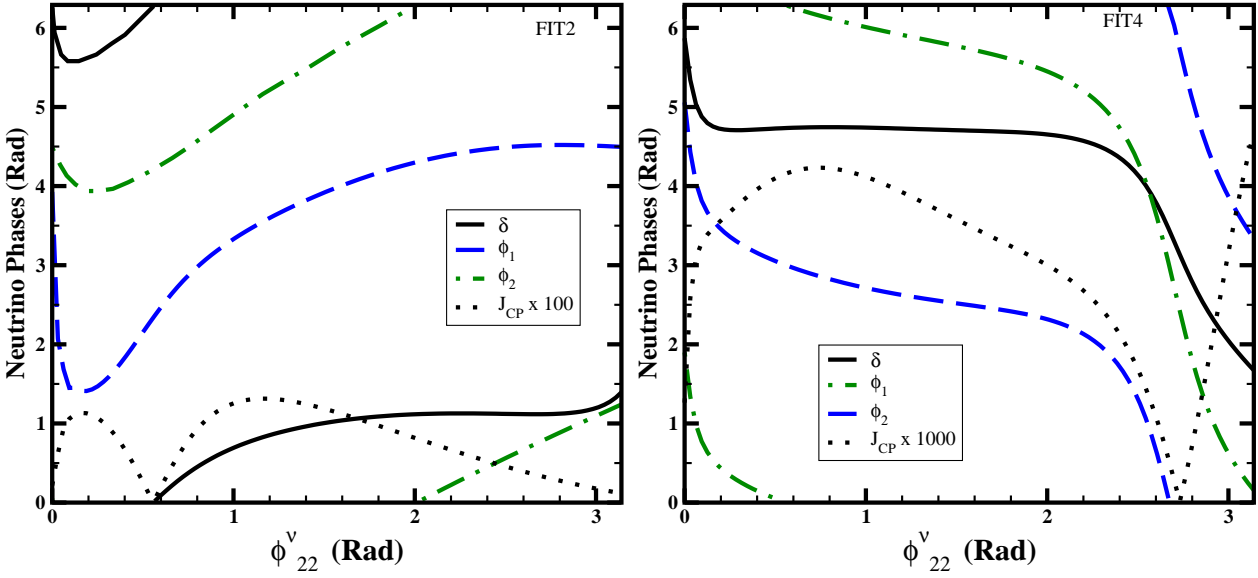

\hspace*{-.35in}
\begin{minipage}{7in}
\epsfig{file=f2-phases.eps,height=3.0in}
\epsfig{file=f4-phases.eps,height=3.0in} 
\hfill
\end{minipage}
\vspace*{0.5 cm}
\caption{\it The neutrino phases and values of the CP-violating Jarlskog invariant $J_{CP}$ as functions of the phase $\phi_{22}^\nu \equiv Arg(a_{22}^\nu)$ 
for Fit~2 and Fit~4, respectively.}
\label{fig:fits-phases}
\end{figure}

\begin{table}[!ht]
\begin{center}
\begin{tabular}{|r|r|r|r|}
\hline   
Case & $\epsilon$ & $\sigma$ & $n_1$\\ \hline \hline
Fit 2 &0.2&19/2&0 \\ \hline
Fit 4 &0.19&0.5&0 \\ \hline \hline
\end{tabular}
\end{center}
\caption{\it Parameter values obtained for Fits 2 and 4 of \cite{K-SU5},
corresponding to the textures of eqns. (9.2, 9.3, 9.4) of that paper.
The values are specified at the GUT scale, the $n_i$ stand for the flavour charges of the
right-handed neutrinos and $\sigma$ is the flavour charge of the singlet 
field that, through
its vev, generates the right-handed neutrino mass hierarchies.}
\label{points}
\end{table}

\begin{table}[!ht]
\begin{center}
\begin{tabular}{|l||l|}
\hline
& Fit 2  of \cite{K-SU5}  \\ \hline \hline
Charged leptons &
$a^e_{12}= 0.55, a^e_{21}= -0.6,
a^e_{22}= 2,  a^e_{23} = -0.7e^{I \pi/4}, 
 a^e_{32}=1.5, a^e_{33} = 1.4$ \\
\hline
$m^\nu_D$ &
$a^\nu_{12} =.9,  
 a^\nu_{22}=2 e^{i \pi/2},   
a^\nu_{23} = 0.5 $ \\
\hline
$M_{RR}$
& $a^N_{11} = 1.1, a^N_{12} = 1.2, a^N_{13} = 1.2, a^N_{22} = 2, a^N_{33} =0.45$ \\
\hline \hline
\end{tabular}
\end{center}
\caption{\it Choice of coefficients to reproduce the fermion observables for Fit 2 of~\cite{K-SU5}.
Coefficients not appearing in the Table are set to unity.}
\label{var1}
\end{table}

\begin{table}[!ht]
\begin{center}
\begin{tabular}{|l||l|}
\hline
& Fit 4  of \cite{K-SU5} \\ \hline
Charged leptons &
$a^e_{13}= 0.8, 
a^e_{22}= -2,  
a^e_{23} = 0.8 e^{i \pi/5}, a^e_{33}=1.1$, \\
\hline
$m^\nu_D$ &
$a^\nu_{13} =0.5,   a^\nu_{22}= 0.7e^{i \pi/4},
a^\nu_{32} =1.9,   a^\nu_{33}= 0.7$ \\
\hline
$M_{RR}$ &
$a^N_{22} =2, a^N_{23} =0.65,a^N_{33} =1.4$ \\
\hline 
\end{tabular}
\end{center}
\caption{\it Choice of coefficients to reproduce the fermion data for Fit 4 of~\cite{K-SU5}.
Coefficients not appearing in the table are set to unity.}
\label{var2}
\end{table}


\begin{table}
\begin{tabular}{|l|l|l|}
\hline

Observables & Series Expansions & Numerical Values \\ \hline
\hline

$m_{\nu_3}$ (eV)  & 
$-0.055 +0.0071\epsilon^{-3/2}+0.0084\epsilon^{-5/4} $  
& 0.05(-0.055)  \\ \hline

$m_{\nu_1}/m_{\nu_3}$  & 
$2.21\epsilon^{2}+0.12\epsilon^{5/2} - 0.051\epsilon^{11/4}$  
& 0.16(0.09)  \\ \hline

$m_{\nu_2}/m_{\nu_3}$ &  
$4.13 \epsilon^{3/2}-0.29\epsilon^{2}-13.82 \epsilon^{9/4}$ 
&0.20(0.37) \\ \hline
\hline

$M_1/M_3$ &  
$0.82\epsilon-0.98\epsilon^{5/4}+0.55\epsilon^{3/2}$ 
& $0.06(0.16)$  \\ \hline

$M_{2}/M_3$  & 
$0.63 \epsilon^{1/4}-1.49\epsilon^{1/2} +1.73\epsilon^{3/4}$ 
&$0.12(0.42)$  \\ \hline
\hline
$m_{\nu_{ee}}$ (eV) 
&$0.017\epsilon^{4/3}+0.04 \epsilon^{5/3}-0.85\epsilon^{2}$ 
& 0.0042(0.002) \\ \hline
\hline
$\theta_{23}$ 
& $1.08+0.17\epsilon^{3/4} -0.012 \epsilon^{4/3}$ 
& 0.91(1.13) \\ \hline

$\theta_{12}$  
& $0.153\epsilon^{1/4}-0.10\epsilon^{2/3}+0.066 \epsilon^{3/4}$ 
& 0.56(0.10) \\ \hline

$\theta_{13}$  
& $0.21 \epsilon^{2/3} +0.25 \epsilon+ 0.0004 \epsilon^{4/3}$ 
& 0.21(0.07) \\ \hline

$\delta$ 
& $-.45+0.059 \epsilon^{1/3}-0.31 \epsilon^{5/12}$ 
& 0.43(-0.45) \\ \hline
\hline
$J_{CP}$ 
& $0.0029 \epsilon^{11/12}+0.003\epsilon^{5/4}+0.0009 \epsilon^{17/12}$ 
& 0.0088(0.00066) \\ \hline
\hline
$\phi_1$
& $2.13+0.23 \epsilon^{1/3}-0.63\epsilon^{5/12}$ 
& 2.94 (2.13) \\ \hline

$\phi_2$
& $2.04+0.23 \epsilon^{1/3}-1.25 \epsilon^{5/12}$ 
& 4.6 (2.03) \\ \hline
\hline

$\epsilon_1$ 
& $0.044\epsilon^{3/2}-0.079\epsilon^{7/4}-0.001\epsilon^{9/4})$ 
& 0.00046(0.0039) \\ \hline

$\epsilon_2$ 
& $0.00046\epsilon^{3/4}+0.0085\epsilon+0.07\epsilon^{5/4}$ 
& 0.0007(0.00014) \\ \hline

$\epsilon_3$ 
& $0.00005\epsilon-0.06\epsilon^{5/4}+0.2\epsilon^{3/2}$ 
& 0.0003(0.000001) \\ \hline
\multicolumn{1}{|l|}
 {${\eta}_B\times \Delta$ }&
 \multicolumn{1}{|c|}{$ - $ } &
 \multicolumn{1}{|l|}{
 $1.18 \times 10^{-7}(3.11\times 10^{-7})$ }
\\ \hline \hline 

$\frac{16 \pi^2}{\kappa}\times m^2_{\tilde{L}_{12}}$ 
&$0.11\epsilon^{2/3}+0.023\epsilon^{11/12}
-2.64 \epsilon$ 
&0.44(0.037) \\ \hline

 \multicolumn{1}{|l|}
 {$BR(\mu\rightarrow e \gamma)$ 
 }&
 \multicolumn{1}{|c|}{$ - $ } &
 \multicolumn{1}{|l|}{
$8.5\times 10^{-12}(6.1 \times 10^{-14})$ 
}
\\ \hline \hline 

$\frac{16 \pi^2}{\kappa}\times m^2_{\tilde{L}_{13}}$ 
&$0.057\epsilon^{2/3}+0.012\epsilon^{11/12}-
0.58 \epsilon$ 
&0.11(0.02) \\ \hline

 \multicolumn{1}{|l|}
 {$BR(\tau\rightarrow e \gamma)$ 
 }&
 \multicolumn{1}{|c|}{$ - $ } &
 \multicolumn{1}{|l|}{
$5.95 \times 10^{-13}(1.74 \times 10^{-14})$ 
}
\\ \hline \hline 

$\frac{16 \pi^2}{\kappa}\times m^2_{\tilde{L}_{23}}$ 
&$0.24+0.05\epsilon^{1/4}+0.002 \epsilon^{1/2}$ 
&0.34 (0.24) \\ \hline

 \multicolumn{1}{|l|}
 {$BR(\tau \rightarrow \mu \gamma)$ 
 }&
 \multicolumn{1}{|c|}{$ - $ } &
 \multicolumn{1}{|l|}{
$5.1\times 10^{-12}(2.6 \times 10^{-12})$ 
}
\\ \hline \hline 

\end{tabular}
\caption{\it Values of observables predicted by Fit 2 of \cite{K-SU5},
for $\tan\beta = 2.04$.
The rate of convergence of each expansion  in $\epsilon$ can be judged
from the relative magnitudes of the expansion coefficients and
by comparing the exact numerical values of the observables with those
obtained by keeping only the first term
in each expansion (in parenthesis).
}
\label{Fit2-obs}
\end{table}


\vspace*{0.8 cm}
\begin{table}
\begin{tabular}{|l|l|l|}
\hline

Observables & Series Expansions & Numerical Values \\ \hline
\hline
$m_{\nu_3}$ (eV)  & 
$-0.054 \epsilon+0.024\epsilon^{-1/2} +0.02 \epsilon^{-3/8}$  
& 0.05(-0.054)  \\ \hline

$m_{\nu_1}/m_{\nu_3}$ & 
$0.26 \epsilon^{3/2}+0.50\epsilon^{7/4} + 0.34 \epsilon^{15/8} $  
& 0.069(0.022) \\ \hline

$m_{\nu_2/m_{\nu_3}}$  & 
$1.04\epsilon^{1/2}-1.94 \epsilon^{3/4} +1.24\epsilon^{7/8}$ 
& 0.21(0.45)  \\ \hline
\hline
$M_{1}/M_3$  & 
$0.046 -0.04\epsilon^{1/8} -0.003\epsilon^{1/4}$ 
&$0.034( 0.046)$  \\ \hline

$M_2/M_3$ &  
$0.95- 1.64\epsilon^{1/8}+1.41\epsilon^{1/4}$ 
& $0.15( 0.95)$  \\ \hline
\hline

$m_{\nu_{ee}}$(eV) & 
$0.026 \epsilon^{7/4} +0.15 \epsilon^{2}+0.05\epsilon^{17/8}$ 
& $0.0027(0.0014)$ \\ \hline
\hline
$\theta_{23}$ 
& $0.90 - 0.027 \epsilon^{3/8}-0.53\epsilon^{1/2}$ 
& 0.87(0.66)\\ \hline

$\theta_{12}$  
& $0.5\epsilon^{1/2}+0.94\epsilon^{3/4}+0.026\epsilon^{7/8}$ 
& 0.53(0.22) \\ \hline

$\theta_{13}$  
& $0.54\epsilon-0.27\epsilon^{11/8}-0.88\epsilon^{3/2}$ 
& 0.04(0.10) \\ \hline

$\delta$ 
& $2.56-0.025 \epsilon^{1/4}+0.03\epsilon^{3/8}$ 
& 4.87(2.56) \\ \hline
\hline
$J_{CP}$ 
& $0.036\epsilon^{3/2}+0.068\epsilon^{7/4}-0.017\epsilon^{15/8}$ 
& 0.004(0.003) \\ \hline
\hline
$\phi_1$ 
& $5.19-0.074 \epsilon^{1/4}+0.096\epsilon^{3/8}$ 
& 6.17(5.20) \\ \hline

$\phi_2$ 
& $2.056+0.025 \epsilon^{1/4}+0.096\epsilon^{3/8}$ 
& 2.91(2.06) \\ \hline
\hline

$\epsilon_1$ 
& $0.000085\epsilon^{3/8}+0.00017\epsilon^{1/2}-0.00013\epsilon^{5/8}$  
& $0.00087(0.000046)$\\ \hline

$\epsilon_2$ 
& $0.0013\epsilon^{3/8}-0.0012\epsilon^{1/2}+0.00053\epsilon^{5/8}$  
& $0.0012(0.00071)$\\ \hline

$\epsilon_3$ 
& $0.0014\epsilon^{3/8}+0.0013\epsilon^{1/2}+0.0006\epsilon^{5/8}$  
& $0.0002(0.00074)$\\ \hline

\multicolumn{1}{|l|}
 {${\eta}_B\times \Delta$ }&
 \multicolumn{1}{|c|}{$ - $ } &
 \multicolumn{1}{|l|}{
 $1.62 \times 10^{-7}(1.03\times 10^{-7})$ }
\\ \hline 
\hline 

$\frac{16 \pi^2}{\kappa}\times m^2_{\tilde{L}_{12}}$ 
&$0.17 \epsilon^{5/8}+0.00024 \epsilon^{3/4}-0.046 \epsilon^{7/8}$ 
&$0.05(0.06)$ \\ \hline

 \multicolumn{1}{|l|}
 {$BR(\mu\rightarrow e \gamma)$ 
 }&
 \multicolumn{1}{|c|}{$ - $ } &
 \multicolumn{1}{|l|}{
$7.8\times 10^{-13}(1.04\times 10^{-12})$ 
}
\\ \hline \hline 

$\frac{16 \pi^2}{\kappa}\times m^2_{\tilde{L}_{13}}$ 
&$0.32\epsilon^{5/8}+0.0066 \epsilon^{3/4}-0.062\epsilon^{7/8}$ 
&$0.16(0.12)$ \\ \hline

 \multicolumn{1}{|l|}
 {$BR(\tau \rightarrow e \gamma)$ 
 }&
 \multicolumn{1}{|c|}{$ - $ } &
 \multicolumn{1}{|l|}{
$7.9\times 10^{-12}(3.9\times 10^{-12})$}
\\ \hline \hline 

$\frac{16 \pi^2}{\kappa}\times m^2_{\tilde{L}_{23}}$ 
&$0.24+0.12 \epsilon^{1/8}+0.035 \epsilon^{1/4} $ 
&$0.48(0.24)$ \\ \hline

 \multicolumn{1}{|l|}
 {$BR(\tau\rightarrow \mu \gamma)$ 
 }&
 \multicolumn{1}{|c|}{$ - $ } &
 \multicolumn{1}{|l|}{
$6.6\times 10^{-11}(1.7\times 10^{-11})$ 

}
\\ \hline \hline 

\end{tabular}
\caption{\it Values of observables predicted by Fit 4 of \cite{K-SU5},
for $\tan\beta = 5.15$. The level of convergence of each expansion  in $\epsilon$ can be judged
from the relative magnitudes of the expansion coefficients and
by comparing the exact numerical values of the observables with that
obtained by keeping only the three first terms in each expansion
(shown in parenthesis).} 
\label{Fit4-obs}
\end{table}

As in the first model example, the changes induced by 
altering the numerical values of the coefficients $a_{ij}$ affect the coefficients in the 
expansions of the mass matrices, but
not the powers of $\epsilon$. However, when varying the absolute values
and the phases of the coefficients $a^e_{ij}$,
the numerical values of the observables are generally affected.
We note the following points in connection with these two examples.

(i) In both fits, the dominant terms in the
expansions generally agree only in order of magnitude
with the exact numerical values, for the values of  $\epsilon$ 
required to reproduce the correct fermion mass hierarchies.
The leading term in many expansions is of order
$z \equiv \epsilon^{1/8}$, and yields good numerical results only when
$z \sim 0.5-0.75$ (i.e., $\epsilon \sim 0.05-0.1$). This slow convergence leads to
the numerical instabilities that are evident in many observables, for example in $J_{CP}$, as shown in Fig.~\ref{fig:fits-jcp}. These instabilities may be introduced 
by combinations of the following:
(a) the powers of $\epsilon$ in the series are often very close to each other
(e.g., $a \epsilon^{3/8}+ b \epsilon^{4/8}
+ c \epsilon^{5/8}$), and
(b) the relative signs of the terms in the expansions may
change with the choice of $a_{ij}$.

(ii) The model predictions for $\theta_{23}$ is ${\cal O}(1)$, and that for $\theta_{12}$ is
suppressed only by ${\cal O}(\epsilon^{1/4})$. As seen in Fig.~\ref{fig:fits}, the numerical
values of these angles are reasonable for $\phi^\nu_{22}$ in Fit~2, in which case
the prediction for $\theta_{13}$ is close to the present experimental upper limit, even
though parametrically it is suppressed by ${\cal O}(\epsilon^{2/3})$. In the case of Fit~4,
the values of $\theta_{23,12}$ are reasonable for $0.3 < \phi^\nu_{22} < 1$, but
$\theta_{13}$ is always considerably below the present experimental upper limit.

(iii) The complex phases $\delta$ and $\phi_{1,2}$ are ${\cal O}(1)$ in both fits, but
vary widely with $\phi^\nu_{22}$, as seen in Fig.~\ref{fig:fits-phases}.

(iv) Observable CP-violation is to be expected in neutrino oscillations, although
the Jarlskog invariant $J_{CP}$ is significantly
smaller than in the case of large $\tan\beta$, as also seen in Fig.~\ref{fig:fits-phases},
particularly for Fit~4. In this later case, there is a rather good agreement
between the numerical value and the dominant terms in the expansion.

(v) Both Fit 2 and Fit 4  may accomodate successful leptogenesis,
and also lead to lepton flavour violation that is
close to the present experimental upper limits. Values  of $\eta_b$ in the range of eq.~(\ref{eq:etarange}) would be obtained with 
the values  presented in Table~\ref{Fit2-obs} and Table~\ref{Fit4-obs}  
if the dilution factor
$\Delta\sim 200$. As we discussed in the previous 
subsection, a global scaling factor $f_s$ for $m_D^\nu$ and the corresponding
rescaling of  $M_{RR}$
would  imply a modification of the leptogenesis (and lepton-flavour violation)
predictions. We find that 
a factor $f_s\sim 0.06-0.07$ would predict $\eta_B$ in the 
range of eq.~(\ref{eq:etarange}) with  
$\Delta\sim 1$ and $M_3 \sim 10^{12}$~GeV. 
However, this 
would imply a reduction of about two orders of magnitude in the predictions 
for charged-lepton-flavor violation.

(vi) The numerical results shown in Fig.~\ref{fig:fits} further demonstrate 
the inter-correlations between the
different physical parameters,
indicating how the experimental limits on neutrino masses
constrain the allowed range for the angles, phases, and $J_{CP}$.
We see that, even for the same GUT group and range  of $\tan\beta$, 
the relations between the observables display
certain differences, reflecting among others the
role of the coefficients in obtaining viable solutions in these schemes.

\subsection{Flipped SU(5) Model}

In the case of the 
flipped SU(5) GUT model, the fields $Q_{i},d_{i}^{c}$ and $\nu_{i}^{c}$ of each family belong to a 
{\tt 10} representation of SU(5), the $u_{i}^{c}$ and $L_{i}$ belong to {\tt $\overline{5}$}  representations, and the $e_{i}^{c}$ fields belong to
singlet representations of SU(5). Thus the correlations between the U(1) charges of the different matter fields are different from those in conventional SU(5).
These particle assignments imply a symmetric down-quark mass matrix, whereas
the structure of the up-quark mass matrix depends on the charges of the
right-handed quarks. However, as these are the same as the charges of the
left-handed leptons, the up-quark mass matrix is constrained by the need to
generate large mixing for atmospheric neutrinos. 

The simplest example that matches the charged-fermion mass hierarchies
(adjusting coefficients so as to match the experimental value of $V_{cb}$) is \cite{LR}
\begin{equation}
M_{down}\propto \left( 
\begin{array}{ccc}
\bar{\epsilon}^{8} & \bar{\epsilon}^{3} & \bar{\epsilon}^{4} \\ 
\bar{\epsilon}^{3} & \bar{\epsilon}^{2} & \bar{\epsilon} \\ 
\bar{\epsilon}^{4} & \bar{\epsilon} & 1
\end{array}
\right),
M_{up}\propto \left( 
\begin{array}{ccc}
{\epsilon }^{|-4+y|} & {\epsilon }^{4} & {\epsilon }^{4} \\ 
{\epsilon }^{|1+y|} & {\epsilon } & \epsilon \\ 
{\epsilon }^{|y|} & 1 & 1
\end{array}
\right),
\end{equation}
\begin{equation}
M_{\ell }\propto \left( 
\begin{array}{ccc}
\bar{\epsilon}^{|a+y|} & \bar{\epsilon}^{|b+y|} & \bar{\epsilon}^{|y|} \\ 
\bar{\epsilon}^{|a|} & \bar{\epsilon}^{|b|} & 1 \\ 
\bar{\epsilon}^{|a|} & \bar{\epsilon}^{|b|} & 1
\end{array}
\right), 
M_{\nu}^D \propto \left( 
\begin{array}{ccc}
{\epsilon }^{|-4+y|} & {\epsilon }^{|1+y|} & {\epsilon }^{|y|} \\ 
{\epsilon }^4 & {\epsilon } & 1 \\ 
{\epsilon }^4 & {\epsilon } & 1 
\end{array}
\right).
\end{equation}
Once again, the form of the heavy Majorana mass matix depends on the charge
of the field $\Sigma$. For instance, for
$\sigma = 0$, it will be similar in structure
to the down-quark mass matrix. The
contribution from the up-quark sector to $V_{cb}$ is generically small in this model,
leading to  $V_{cb}\simeq \sqrt{m_{s}/m_{b}.}$ 
(indeed, for the up-type quark hierarchies, it turns out that 
$\bar{\epsilon} \approx 0.2, ~\epsilon \sim \bar{\epsilon}^4$ 
and $|y| \sim 2$). This is too large, and requires a significant 
adjustment of the ${\cal O}(1)$ coefficients.

However, the problem with the large value 
of $V_{cb}$ can be avoided by combining flipped SU(5)
with a non-Abelian flavour group, or by 
adding a second singlet field with different transformation
properties under the flavour group.
In that way, one could obtain solutions similar to those of~\cite{nonab-KR}, with 
\begin{equation}
M_{down}\propto \left( 
\begin{array}{ccc}
\bar{\epsilon}^{8} & \bar{\epsilon}^{3} & \bar{\epsilon}^{4} \\ 
\bar{\epsilon}^{3} & \bar{\epsilon}^{2} & \bar{\epsilon}^2 \\ 
 \bar{\epsilon}^{4}   & \bar{\epsilon}^2 & 1
\end{array}
\right).
\end{equation}
However, even after overcoming this obstacle, it is very difficult to obtain naturally viable solutions: if $\epsilon \sim \bar{\epsilon}^4$, the Dirac-mass hierarchies turn out to be too large to generate interesting solutions.
Of course, the fact that the see-saw mechanism is hard to implement does not mean
that an appropriate $m_{eff}$ could not be generated by alternative mechanisms.
However, the symmetries require that
\[
m_{eff}\propto \left( 
\begin{array}{ccc}
\tilde{\epsilon}^{|2y|} & \tilde{\epsilon}^{|y|} & \tilde{\epsilon}^{|y|} \\ 
\tilde{\epsilon}^{|y|} & 1 & 1 \\ 
 \tilde{\epsilon}^{|y|}   & 1 & 1
\end{array}
\right)
\]
indicating that, unlike in conventional SU(5), it is difficult to generate
large solar mixing.

As in the case of conventional SU(5), one could either
(i) study the most generic cases
with half-integer charges, requiring the (3,3) flavour charges to be non-zero,
and using the see-saw conditions to obtain large mixing for the solar 
neutrinos, or
(ii) add further fields and flavour groups.
However, any departure from the simple structure described so far introduces
significant model dependence, and results in a loss of predictivity, unless
this additional structure is predicted (and constrained) by concrete theoretical
considerations. In the subsection that follows, we discuss what may be the most natural way to proceed, if one believes in the existence of an underlying string
theory.

\subsection{String-Derived Flipped SU(5) Models}

So far, we have been discussing models based 
on a single U(1) flavour symmetry, and with only one
field $\theta$ whose vev provides a single expansion parameter in the mass matrices.
However, in realistic models this need not be the case. On the contrary, one may expect several
fields to be involved in the generation of mass terms, 
while additional flavour symmetries may be relevant.
While this seems to limit predictivity, in models based on an underlying string theory,
the string symmetries (translated into selection rules) impose strong constraints on the mass 
and mixing matrices. 

In the previous sections we observed that there 
is significant model-dependence in the results,
and that the values of certain coefficients must be in specific ranges in order to give viable
solutions. Optimally, we would like to understand these
values from prior principles. In a realistic model, they could be related to the vacuum expectation 
values of fields
that are constrained, for example, by considerations on flat directions.

In specific string-derived  models, due to the string selection rules,
several texture zeros are generated in the matrices, and these could in principle
lead to strong constraints on certain observables. Indeed, the constraints are so strong
that one could expect that it might be difficult to generate the required entries that fully
reproduce the observed fermion patterns. Nevertheless, it was found in \cite{ELLN, ELLN2} 
that it is  possible  to  accommodate all data 
in a natural and generic way within a flipped
SU(5) $\times$ U(1) string model~\cite{aehn}.
Relevant aspects of this model are reviewed
in Appendix II: the theory contains 
many singlet fields, and the mass matrices depend on the subset of them
that get non-zero vev's, i.e., on the 
choice of flat directions in the effective potential.

The flat directions and the quark masses and mixings for this model
have been studied in \cite{ELLN}, where it was found that

\beq
M_D = \left (
\begin{array}{ccc}
0 & \Delta_2 \Delta_3^2 \bar{\Phi}_{23} & 
\Delta_5 \Delta_3 \bar{\phi}_{3} \\ 
\Delta_2 \Delta_3 \bar{\Phi}_{23} & 
(\bar{\phi}_3^2 + \bar{\phi}_4^2) &
\Delta_2 \Delta_5 \bar{\phi}_{4} \\ 
\Delta_5 \Delta_3 \bar{\phi}_{3} & 
\Delta_2 \Delta_5 \bar{\phi}_{4} & 1
\end{array}
\right ),
~~~M_U = \left (
\begin{array}{ccc}
0 & 0 & \Delta_5 \Delta_3 \bar{\phi}_{3} \\ 
0 & \bar{\phi}_4 & \Delta_2 \Delta_5 \bar{\phi}_{4} \\ 
0 &  \Delta_2 \Delta_5 & 1
\end{array}
\right ).
\eeq
As already remarked, the relevant field
definitions are given in Appendix II. For instance,
$\Delta_2 \Delta_5$ is a combination of 
hidden-sector fields that transform as
sextets under SO(6).

The lepton mass matrices were studied in \cite{ELLN2},
where it was shown that:
\beq
m_{\ell}  \propto \left (
\begin{array}{ccc}
{\bar \phi}_4^2 & \Delta_2 \Delta_5 {\bar \phi}^2_3 & 0 \\
0 & {\bar \phi}^2_{3} & 0 \\
0 & 0 & 1
\end{array}
\right ) \equiv \left (
\begin{array}{ccc}
f^2& x u^2&0\\ 
0 & u^2 & 0\\
0 & 0 & 1
\end{array}
\right ),
\label{eq:fllep}
\eeq

\beq
m_{\nu}^D \propto 
\left (
\begin{array}{ccc}
\Delta_2 \Delta_5 \bar{\phi}_4 & 1 & 0 \\
\bar{\phi}_4 & \Delta_2 \Delta_5 & 0 \\
0 & 0 & F_1
\end{array}
\right )\equiv \left (
\begin{array}{ccc}
f x& 1 &0\\ 
f & x & 0\\
0 & 0 & F_1
\end{array}
\right ),
\label{eq:flnu}
\eeq
\beq
M_{\nu_R} \propto
\left (
\begin{array}{ccc}
\bar{F}_5 \bar{F}_5  \bar{\phi}_4 \phi_3 &
\bar{F}_5 \bar{F}_5  \Delta_2 \Delta_5 \phi_3 & 0 \\
\bar{F}_5 \bar{F}_5  \Delta_2 \Delta_5 \phi_3 &
0 & \bar{F}_5 \bar{\Phi}_{31}\Phi_{31} \bar{\phi}_4 \phi_2 \\
0 & \bar{F}_5 \bar{\Phi}_{31}\Phi_{31} \bar{\phi}_4 \phi_2 
& \Delta_2 \Delta_5 \bar{\Phi}_{23} T_2 T_5
\end{array}
\right) \equiv \left (
\begin{array}{ccc}
f r y^2 & 2 r x y^2 &0\\ 
2 r x y^2& 0 & b f y\\
0 & b f y  & c t x
\end{array}
\right ),
\label{eq:flmr}
\eeq
where 
\beq
y \equiv \bar{F}_5, r \equiv \phi_3, b \equiv \bar{\Phi}_{31} \Phi_{31} \phi_2, c \equiv \bar{\Phi}_{23},\bar{\phi}_3 \equiv u .
\eeq

Note that these matrices are given in the flipped SU(5) field basis. However it is easy to pass to the weak interaction eigenstates by an 
appropriate rotation. For instance, 
the weak-interaction eigenstates for the light neutrinos have the
following assignments:
\bea
\nu_{e} \rightarrow \bar{f}_5 - {\cal O}(\Delta_2 \Delta_5) {\bar f}_2,
\;\;\;\;
\nu_{\mu} \rightarrow \bar{f}_2 + {\cal O}(\Delta_2 \Delta_5){\bar f}_5
\;\;\;\;
\nu_{\tau} \rightarrow \bar{f}_1 ,
\label{assignneutrinos}
\eea
while the flipped SU(5) basis 
$({\bar f}_5, {\bar f}_2, {\bar f}_1)$ 
is related to (\ref{assignneutrinos}) by the rotation
\beq
V^m_{\ell_L} = \left (
\begin{array}{ccc}
1-\frac{1}{2} (\Delta_2 \Delta_5)^2 & \Delta_2 \Delta_5 & 0 \\
-\Delta_2 \Delta_5  &  1-\frac{1}{2} (\Delta_2 \Delta_5)^2 & 0 \\
0 & 0 & 1
\end{array}
\right ).
\eeq

The forms of the mass matrices depend
on the various field vev's. 
The analysis of quark masses pointed towards
$x=\Delta_2 \Delta_5 = {\cal O} (1)$ (and potentially large solar
mixing, already from the charged-lepton sector),
as well as a rather suppressed value of $f=\bar{\phi}_4 \ll 1$
(since $V_{cb}\approx\Delta_2 \Delta_5 \bar{\phi}_4$).
Moreover, from the analysis of flat directions~\cite{ELLN},
it was concluded that $\bar{\Phi}_{31} \bar{\Phi}_{23} = {\cal O} (1)$ 
is large. In addition,
the flatness conditions~\cite{ELLN} 
relate $\bar{\Phi}_{31},\Phi_{31}$ and
$\phi_2$, and can be satisfied even if all
the vev's are large, as long as
$\bar{\Phi}_{31} \Phi_{31}$ and
$\bar{\Phi}_{23} \Phi_{23}$ are
not very close to unity.
As for the
decuplets that break the gauge group down to
the Standard Model, we know that the vev's 
should be 
$\approx M_{GUT}/M_{s}$. In weakly-coupled string constructions,
this ratio is $\approx 0.01$. However, the strong-coupling limit of
$M$ theory offers the possibility that the
GUT and the string scales can coincide, in which case the
vev's could be of order unity. 

The $2 \times 2$ form of the charged-lepton mass matrix in fact puts severe constraints on
the fields involved.
On the other hand, flatness conditions and quark
masses do not give any information on the vev of the product
$T_2 T_5$. However, this combination is to some extent
constrained by the
requirements for the light neutrino masses \cite{ELLN2}.
Finally, the field $\phi_3$ is the one for which we seem
to know least and we can consider two extreme possibilities:
a very large value ${\cal O}(1)$ and a very small value.
For large $\phi_3$, if 
$\bar{\phi}_4 \approx \bar{F}_5,F_1$
as would be expected in weak-coupling
unification schemes,
the entries of $m_{eff}$ are all of the same order of magnitude,
and large
$\nu_{\mu}-\nu_{e}$ and 
$\nu_{\mu}-\nu_{\tau}$ mixings are naturally generated via the neutrino mixing
matrix.


Charged-lepton mass hierarchies pose severe constraints on the model parameters.
From (\ref{eq:fllep}) we get
\beq
m_e/m_\mu\sim {f^2\over {u^2(1+x^2)}},
\eeq
while $f$ and $x$ are also constrained from 
the quark masses, by
\bea
m_c/m_t&\sim& f\times \rm{\cal O}(1),\\
V_{cb}&\sim& x f \sim 0.044.\\ \nonumber
\eea  

Therefore, if  $f$ is $\rm{\cal O}(0.01)$, $x$ can be 
$\rm{ \cal O}(1)$ while the charged leptons mass ratio can be satisfied with 
$u$ of $\rm{ \cal O}(0.1)$.  

The charged-lepton mixing matrix $V_e$ is obtained by the diagonalization of 
$m_{\ell}$ in the weak-interaction basis:
\beq
V_e^T (V^m_{\ell_L})^T m_{\ell} m_{\ell}^\dagger (V^m_{\ell_L})^* V_e^*= 
Diag(m_e^2,m_\mu^2,m_\tau^2).
\eeq
We parametrize the product of these two rotations by defining:
\beq
Q_e=V_e V^m_{\ell_L} \sim \left (
\begin{array}{ccc}
1/N& g/N & 0 \\
-g/N &  1/N & 0 \\
0 & 0 & 1
\end{array}
\right ),
\eeq 
where $g$ is a mixing parameter to be determined by the neutrino mass matrix,
and $N$ is a normalization factor for the sector 1-2: $N=1/(1+|g|^2)^{1/2}$. 

\begin{figure}[!ht]
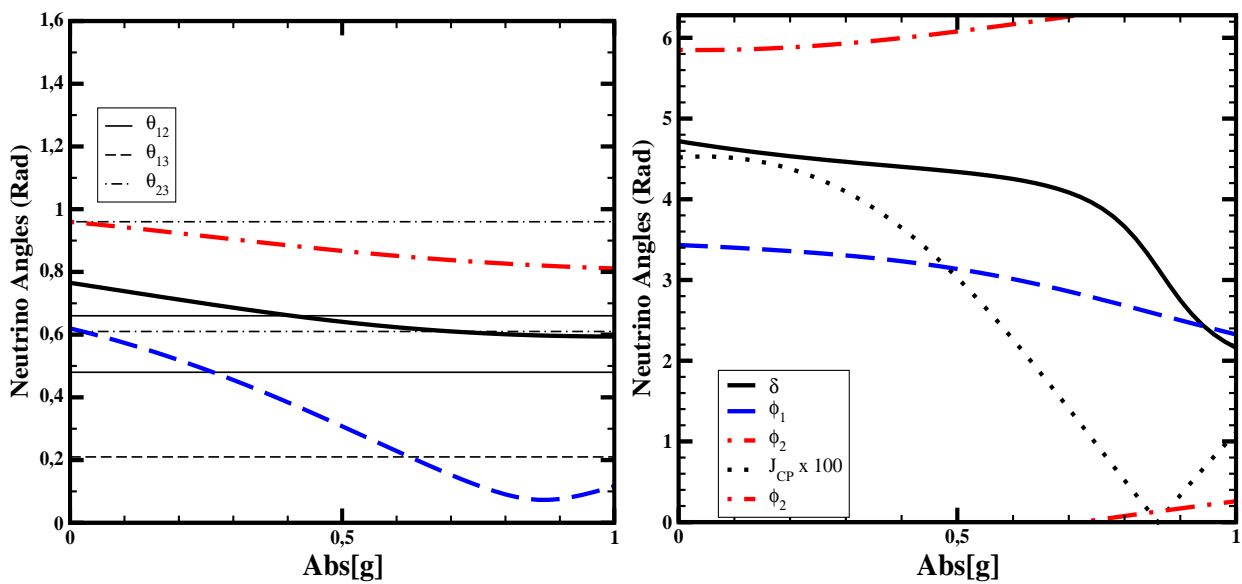

\begin{center}
\epsfig{file=flip_angles.eps,height=3.0in,width=3.2in}
\epsfig{file=flip_phases.eps,height=3.0in,width=3.2in}
\end{center}
\caption{\it The neutrino mixing angles (left) and the phases and 
 value of the CP-violating Jarlskog invariant $J_{CP}$ (right), as functions of the absolute value of $g$.
The experimental limits on the three neutrino angles
are denoted by horizontal lines.}
\label{fig:flip_angles}
\end{figure}

Passing finally to the neutrino mass matrix, we see the following.
In the basis where the charged-lepton masses are diagonal, ({\it i.e.} 
$m_\nu^D\rightarrow Q_e^T\cdot m_\nu^D$), the vev's for the remaining fields $m_\nu^D$ and $M_{\nu_R}$ enter in 
$m_{eff}$ as combinations:
\beq
m_{eff}\propto
 \left (
\begin{array}{ccc}
(G_{11}+\rm{\bf p}\cdot H_{11})/N^2&(G_{12}+ \rm{\bf p} \cdot H_{12})/N^2&
 \rm{\bf q}\cdot (1+g x-2 x^2)/N\\

(G_{12}+\rm{\bf p}\cdot H_{12})/N^2&
(G_{22}+\rm{\bf p}\cdot H_{22})/N^2&
-\rm{\bf q} \cdot (x-g(1-2x^2))/N\\

\rm{\bf q}\cdot (1+g x-2 x^2)/N&
-\rm{\bf q} \cdot (x-g(1-2x^2))/N&
4 \rm{\bf q}^2\cdot x^2\\

\end{array}
\right).
\eeq

Here, {\bf p, q} are combinations of vev's defined as:
\beq
\rm{\bf p}={c\cdot r\cdot t \over b^2 \cdot f^3}, 
\rm{\bf q}={y\cdot f_1 \cdot r \over b \cdot f^2},
\label{eq:pq}
\eeq
while the functions  $G_{ij}(x,g)$, $H_{ij}(x,g)$,  are of  ${\cal O} (1)$:
\bea
G_{11}(x,g)&=& g^2 - 2 g x + x^2,\nonumber \\
G_{12}(x,g)&=& x - g^2 x + g (-1 + x^2),\nonumber \\
G_{22}(x,g)&=&1 + 2 g x + g^2 x^2,\nonumber \\
H_{11}(x,g)&=&3 g^2 x^3 + x (-1 + 4 x^2) - 2 g x (x + 2 x^3),\nonumber \\
H_{12}(x,g)&=&g x (-1 + x^2) + x (x + 2 x^3) - 
      g^2 x (x + 2 x^3) ,\nonumber \\
H_{22}(x,g)&=&3 x^3 + g^2 x (-1 + 4 x^2) + 2 g x (x + 2 x^3).
\eea

The phase of $\rm{\bf q}$ can be  eliminated with a rotation
$q_3\cdot m_{eff}\cdot q_3$, where
$q_3=diag(1,1,Arg(q)/2)$. The neutrino data can be fit by choosing suitably the parameters $\rm{\bf p}$,  
$\rm{\bf q}$,  $g$ and $x$. 
The results  presented in Table~\ref{flip1}
correspond to following  choice of parameters:
\beq
x=0.8\cdot e^{0.3 i},  \ \ \ g=0.7 \cdot e^{-0.94 i}, \\\ \rm{\bf p}=0.43\cdot e^{2.2 i}, 
\\\ \rm{\bf q}=0.25,
\label{eq:flpara} 
\eeq
while in Fig.~\ref{fig:flip_angles} we vary the value of $|g|$ while keeping constant the rest 
of the parameters.


To estimate the values of the observables that involve the right-handed neutrinos, we 
need the individual values of the vev's contained in {\bf p, q}. Since, 
b, c, r are $\cal{O}$(1), $t\sim f^3$ and also $y\sim F_1\sim f$. In 
order to provide some numerical values we can make the assignment
$y=F_1=t^{1/3}=|f|=0.01$.
Then, from the values of {\bf p, q} we can get:
\beq
\left|r/b\right|=1/4,\ \ \  \left|c/b\right|=1.84.
\eeq
Despite the phase dependence of $M_{\nu R}$ being rather complex since all the 
vev's contain phases, these appear  on the observables 
$\epsilon_{1,2}$ and BR($\mu \rightarrow e \gamma$) in  the same combination as the phase of {\bf p}. 
For field  vev's chosen as above,
the matrix $M_{\nu_R}$ in (\ref{eq:flmr}) contains two almost 
degenerate eigenvalues $M_{1,2}\sim \left|(M_{\nu_R})_{12}^2+
(M_{\nu_R})_{23}^2\right|^{1/2}$.
The predictions of the string-derived flipped SU(5) model for the neutrino masses,
mixing angles, CP  and charged-lepton flavour violation are shown in Table~\ref{flip1}.

\begin{table}[!ht]
\begin{center}
\begin{tabular}{|l|l|}
\hline
Observables& Numerical Values \\ \hline
\hline 
$m_{\nu_3}$ (eV) & 0.056 \\ \hline 

$m_{\nu_1}/m_{\nu_3}$ & 0.008  \\ 
\hline 

$m_{\nu_2}/m_{\nu_3}$ & 0.17 \\ \hline 
\hline 

$M_{1}/M_3$  & 
$1.4\times 10^{-4}$  \\ \hline

$M_2/M_3$ &  
$0.98$  \\ \hline 
\hline 

$m_{\nu_{ee}}$(eV) & 0.0027 \\ \hline 
\hline 

$\theta_{23}$  
& 0.84 \\ \hline

$\theta_{12}$  
&0.61\\ \hline

$\theta_{13}$ 
&0.15 \\ \hline

$\delta$ & 4.1\\ \hline 
\hline 

$J_{CP}$ &0.014  \\ \hline \hline

$\phi_1$ & 2.86 \\ \hline

$\phi_2$ & 6.26 \\ \hline 
\hline 

$\epsilon_1$ 
& $1.5\times 10^{-7}$ \\ \hline

$\epsilon_2$ 
& $3.1\times 10^{-3}$\\ \hline 

$\epsilon_3$ 
& $3.2\times 10^{-3}$\\ \hline
$\eta_B\times \Delta$ 
& $4.5\times 10^{-8}$\\ \hline 
\hline 

$BR(\mu\rightarrow e \gamma)\times \frac{1}{(tan\beta)^2}$ 
&$ 2.4\times 10^{-10}$ \\
\hline \hline 
\end{tabular}
\end{center}
\caption{\it Values of observables predicted in the string--derived flipped 
SU(5) model for the choice of parameters given in the text.}. 
\label{flip1}
\end{table}

The following are some notable features.

(i) Since the vev's are fixed in the string-derived flipped SU(5) model, there is no
auxiliary expansion parameter, as in the previous SU(5) models. However, it is
interesting to study observables as a function of $g$, as shown in 
Fig.~\ref{fig:flip_angles}.

(ii) We see in Fig.~\ref{fig:flip_angles} that $\theta_{23}$ is large and within the experimental range for all values of $g$,
whereas $\theta_{12}$ lies within the allowed range for $|g| > 0.5$, and $\theta_{13}$ is
acceptably small for $|g| > 0.6$ and may lie close to the experimental upper limit.

(iii) We also see in Fig.~\ref{fig:flip_angles} that $\delta$ and the two low-energy Majorana
phases are expected to be quite large.

(iv) We also see that $J_{CP}$ may be quite large, though there is also the possibility that it might
(be close to) vanish(ing).

(v) We see in Table~\ref{flip1} that leptogenesis is quite possible. Values  
of $\eta_b$ in the range of eq.~(\ref{eq:etarange}) with 
the values  presented in Table~\ref{flip1} 
are consistent with a dilution factor $\Delta\sim 80$. In this case, we 
find that 
a global scaling factor $f_s \sim 0.11$ on  $m_D^\nu$  
would predict $\eta_B$ in the range of eq.~(\ref{eq:etarange}) with  
$\Delta\sim 1$ and $M_3 \sim 10^{12}$~GeV. Such a scaling 
would decrease the prediction for BR($\mu\rightarrow e \gamma$) by two 
orders of magnitude, making it compatible with the  experimental bound 
for a lower range of supersymmetric masses.

(vi) The values of Table~\ref{flip1} correspond to 
$M_3=10^{14}$~GeV. However, these  imply
a BR($\mu \rightarrow e \gamma$) above the experimental bound
(unless the supersymmetry breaking masses become very large), due to
$\cal{O}$(1) entries in the Dirac neutrino mass matrix 
(\ref{eq:flnu}). Clearly, by lowering the heavy
right-handed neutrino masses, and thus also the Dirac neutrino couplings,
the rates are altered  accordingly. For example, $M_3=5 \times 10^{12}$~GeV would yield
BR($\mu \rightarrow e \gamma$)~$\sim 1.4 \cdot 10^{-12}\cdot \tan\beta^2$,
while $\epsilon_1\sim \epsilon_2\sim  1.5 \cdot 10^{-4}$.
The expressions for $\epsilon_{1,2}$ and 
BR($\mu \rightarrow e \gamma$) contain $Y_\nu$ in a basis where 
 $M_{\nu_R}$ is diagonal, so their values have the same phase 
dependence as in the 1-2  sector of $m_{eff}$. Hence, with the choice of 
parameters of (\ref{eq:flpara}), the results presented in 
Table~\ref{flip1}  are independent of the individual phases of 
the parameters included in the definition of {\bf p} in (\ref{eq:pq}).

(vii) On the contrary, due to the lack of mixing
in the third generation in this model,
$\tau$ decays are not observable in this case.

\section{Conclusions}

We have studied in this paper the predictions for CP and  charged-lepton-number violation in
different SU(5) models, including a string version of flipped SU(5).
Because of the inter-correlations between the different neutrino observables,
it was in each case possible to obtain quite specific predictions, which differ significantly
between the models. 

The general indications are that
accessible rates for CP and charged-lepton-number violation are to be expected.
However, due to the different expectations for the couplings and angles in the
various models, the magnitudes of the observables may be quite different,
even between models based on the same group. We note that $\theta_{13}$ may be close to the present
experimental upper limit in some models, but much smaller in others. Also, the predictions
for the Jarlskog invariant $J_{CP}$ vary significantly. Among the most sensitive observables are those
for charged-lepton-flavour violation and
leptogenesis, and cases with a resonant enhancement of the expected lepton asymmetry
for degenerate heavy Majorana neutrinos
are particularly interesting.

It is crucial to keep in mind that the various models sometimes contain
instabilities due to cancellations, to which the results may be sensitive.
In principle, for cases such as the flipped SU(5) string model,
we expect that the coefficients are to a 
good extent predicted by the vev's of the fields. However,
one should not forget that there is a sum of non-renormalisable terms,
implying  that, if the higher-order terms involve fields with high vev's,
they can effectively introduce multiplicative factors, particularly in the
case of many fields. In principle, once some of the observables discussed here
are measured, one may use the correlated data to 
refine the predictions with better knowledge of the possible ranges of coefficients
and the level
of tuning via sub-determinant cancellations. However, such an analysis is
premature ahead of the corresponding measurements, and would go beyond the scope of
this paper.

\section*{Appendix I: GUT Fits 2 and 4 of \cite{K-SU5}}

We give here the matrix parametrizations found in Fits 2 and 4, as used in the text.

{\bf Fit 2}
\begin{eqnarray}
  \label{eq:10}
  Y^u \! &=&\! 
   \left[
    \begin{array}{ccc}
      a^u_{11} \epsilon^{16} & a^u_{12} \epsilon^{6} & a^u_{13} \epsilon^{8} \\
      a^u_{21} \epsilon^{6} & a^u_{22} \epsilon^{4} & a^u_{23} \epsilon^{2} \\
      a^u_{31} \epsilon^{8} & a^u_{32} \epsilon^{2} & a^u_{33}
    \end{array}
  \right]\!,
  \quad \quad \quad
  Y^d \ = \  
   \left[
    \begin{array}{ccc}
      a^d_{11} \epsilon^{31/2} & a^d_{12} \epsilon^{11/2} & a^d_{13} \epsilon^{11/2} \\
      a^d_{21} \epsilon^{11/2} & a^d_{22} \epsilon^{9/2} & a^d_{23} \epsilon^{9/2} \\
      a^d_{31} \epsilon^{15/2} & a^d_{32} \epsilon^{5/2} & a^d_{33} \epsilon^{5/2}
    \end{array}
  \right],
  \nn \\
\label{eq:3}
  Y^e &=& 
   \left[
    \begin{array}{ccc}
      a^e_{11} \epsilon^{46/3} & a^e_{12} \epsilon^{16/3} & a^e_{13} \epsilon^{22/3} \\
      a^e_{21} \epsilon^{16/3} & a^e_{22} \epsilon^{14/3} & a^e_{23} \epsilon^{8/3} \\
      a^e_{31} \epsilon^{16/3} & a^e_{32} \epsilon^{14/3} & a^e_{33} \epsilon^{8/3}
    \end{array}
  \right]\!,
  \
  Y^\nu =
   \left[
    \begin{array}{ccc}
      a^\nu_{11} \epsilon^{|n_1 +5|} & a^\nu_{12} \epsilon^{\frac{41}{8}} & a^\nu_{13} \epsilon^{\frac{11}{2}} \\
      a^\nu_{21} \epsilon^{|n_1 - 5|} & a^\nu_{22} \epsilon^{\frac{39}{8}} & a^\nu_{23} \epsilon^{\frac{9}{2}} \\
      a^\nu_{31} \epsilon^{|n_1 -5|} & a^\nu_{32} \epsilon^{\frac{39}{8}} & a^\nu_{33} \epsilon^{\frac{9}{2}}
    \end{array}
  \right],
  \nn\\
  M_{RR} \!&=&\!
   \left[
    \begin{array}{ccc}
      \epsilon^{|2n_1+\sigma|} & \epsilon^{|1/8 + n_1 + \sigma|} & \epsilon^{|1/2 + n_1+\sigma|} \\
      . & a^N_{22} \epsilon^{|1/4+\sigma|} & \epsilon^{|5/8+\sigma|} \\
      . & . & \epsilon^{|1+\sigma|}
    \end{array}
  \right] \left<\Sigma\right>.
\end{eqnarray}

{\bf Fit 4:}
\begin{eqnarray}
Y^u \! &=&\! 
   \left[
    \begin{array}{ccc}
      a^u_{11} \epsilon^{6} & a^u_{12} \epsilon^{5} & a^u_{13} \epsilon^{3} \\
      a^u_{21} \epsilon^{5} & a^u_{22} \epsilon^{4} & a^u_{23} \epsilon^{2} \\
      a^u_{31} \epsilon^{3} & a^u_{32} \epsilon^{2} & a^u_{33}
    \end{array}
  \right]\!,
  \quad \quad \quad
  Y^d \! = \!  
   \left[
    \begin{array}{ccc}
      a^d_{11} \epsilon^{4} & a^d_{12} \epsilon^{3} & a^d_{13} \epsilon^{3} \\
      a^d_{21} \epsilon^{3} & a^d_{22} \epsilon^{2} & a^d_{23} \epsilon^{2} \\
      a^d_{31} \epsilon & a^d_{32}  & a^d_{33}
    \end{array}
  \right]\epsilon^{|k_d|},\nn\\
 Y^e\!&=&\!
 \left[
    \begin{array}{ccc}
      a^e_{11} \epsilon^{4} & a^e_{12} \epsilon^{3} & a^e_{13} \epsilon \\
      a^e_{21} \epsilon^{3} & a^e_{22} \epsilon^{2} & a^e_{23}  \\
      a^e_{31} \epsilon^{3} & a^e_{32}\epsilon^{2}  & a^e_{33}
    \end{array}
  \right]\epsilon^{|k_d|},
\quad 
 Y^\nu \!=\! 
   \left[
    \begin{array}{ccc}
      a^\nu_{11} \epsilon^{|n_1 + 1|} & a^\nu_{12} \epsilon^{5/8} & a^\nu_{13} \epsilon \\
      a^\nu_{21} \epsilon^{|n_1-3/8|} & a^\nu_{22} \epsilon^{3/8} & a^\nu_{23}  \\
      a^\nu_{31} \epsilon^{|n_1|} & a^\nu_{32} \epsilon^{3/8} & a^\nu_{33} 
    \end{array}
  \right],
 \nn \\
    M_{RR}\! &=&\!
 \left[
    \begin{array}{ccc}
       \epsilon^{|2n_1 + \sigma|} &  \epsilon^{|-3/8 + n_1 + \sigma|} &  \epsilon^{|n_1+\sigma|} \\
       . &  a^N_{22}\epsilon^{|-3/4 + \sigma|} & \epsilon^{|-3/8 + \sigma|} \\
       . & . &\epsilon^{|\sigma|}
    \end{array}
  \right] \left<\Sigma\right>.
\label{eq:31}
\end{eqnarray}

\section*{Appendix II: Flipped SU(5) Particle Assignments}

In this Appendix we tabulate for completeness
the field assignment of the `realistic'
flipped SU(5) string model~\cite{aehn},
as well as the basic conditions used in~\cite{ELLN} to obtain consistent
flatness conditions and acceptable Higgs masses.

\begin{table}[!ht]
\begin{center}
\begin{tabular}{|l||l||l|}
\hline
$F_1(10,\frac{1}{2},-\frac{1}{2},0,0,0)$ &
$\bar{f}_1(\bar{5},-\frac{3}{2},-\frac{1}{2},0,0,0)$ &
$\ell_1^c(1,\frac{5}{2},-\frac{1}{2},0,0,0)$ \\
 
$F_2(10,\frac{1}{2},0,-\frac{1}{2},0,0)$ &
$\bar{f}_2(\bar{5},-\frac{3}{2},0,-\frac{1}{2},0,0)$ &
$\ell_2^c(1,\frac{5}{2},0,-\frac{1}{2},0,0)$ \\
 
$F_3(10,\frac{1}{2},0,0,\frac{1}{2},-\frac{1}{2})$ &
$\bar{f}_3(\bar{5},-\frac{3}{2},0,0,\frac{1}{2},\frac{1}{2})$ &
$\ell_3^c(1,\frac{5}{2},0,0,\frac{1}{2},\frac{1}{2})$ \\
 
$F_4(10,\frac{1}{2},-\frac{1}{2},0,0,0)$ &
$f_4(5,\frac{3}{2},\frac{1}{2},0,0,0)$ &
$\bar\ell_4^c(1,-\frac{5}{2},\frac{1}{2},0,0,0)$ \\
 
$\bar{F}_5(\overline{10},-\frac{1}{2},0,\frac{1}{2},0,0)$ &
$\bar{f}_5(\bar{5},-\frac{3}{2},0,-\frac{1}{2},0,0)$ &
$\ell_5^c(1,\frac{5}{2},0,-\frac{1}{2},0,0)$ \\
\hline
\end{tabular}
\label{table:4}
 
\vspace*{0.5 cm}

\begin{tabular}{|l||l||l|}
\hline
$h_1(5,-1,1,0,0,0)$ & $h_2(5,-1,0,1,0,0)$ & $h_3(5,-1,0,0,1,0)$ \\
$h_{45}(5,-1,-\frac{1}{2},-\frac{1}{2},0,0)$ & &  \\
\hline
\end{tabular}

\vspace*{0.5 cm}

\begin{tabular}{|l||l||l|}
\hline
$\phi_{45}(1,0,\frac{1}{2},\frac{1}{2},1,0) $ &
$\phi_{+}(1,0,\frac{1}{2},-\frac{1}{2},0,1) $ &
$\phi_{-}(1,0,\frac{1}{2},-\frac{1}{2},0,-1) $ \\
$\Phi_{23}(1,0,0,-1,1,0)$ &
$\Phi_{31}(1,0,1,0,-1,0)$  &
$\Phi_{12}(1,0,-1,1,0,0)$ \\
$\phi(1,0,\frac{1}{2},
-\frac{1}{2},0,0)$ &
$\Phi(1,0,0,0,0,0)$ & \\
\hline
\end{tabular}
 
\vspace*{0.5 cm}

\begin{tabular}{|l||l||l|}
\hline
$\Delta_1(0,1,6,0,-\frac{1}{2},\frac{1}{2},0)$ &
$\Delta_2(0,1,6,-\frac{1}{2},0,\frac{1}{2},0)$ &
$\Delta_3(0,1,6,-\frac{1}{2},-\frac{1}{2},0,
\frac{1}{2})$ \\
$\Delta_4(0,1,6,0,-\frac{1}{2},\frac{1}{2},0)$ &
$\Delta_5(0,1,6,\frac{1}{2},0,-\frac{1}{2},0)$ & \\
\hline 
\end{tabular}

\vspace*{0.5 cm}

\begin{tabular}{|l||l||l|}
\hline
$T_1(0,10,1,0,-\frac{1}{2},\frac{1}{2},0)$ &
$T_2(0,10,1,-\frac{1}{2},0,\frac{1}{2},0)$ &
$T_3(0,10,1,-\frac{1}{2},-\frac{1}{2},0,\frac{1}{2})$ \\
$T_4(0,10,1,0,\frac{1}{2},-\frac{1}{2},0)$ &
$T_5(0,10,1,-\frac{1}{2},0,\frac{1}{2},0)$ & \\
\hline
\end{tabular}

\vspace*{0.2 cm}
\end{center}

{\small Table 11: 
{\it The chiral superfields are listed with their
quantum numbers \cite{aehn}.
The $F_i$, $\bar{f}_i$, $\ell_i^c$,
as well as the
$h_i$, $h_{ij}$ fields and the singlets
are listed with their
$ SU(5) \times U(1)' \times U(1)^4$ 
quantum numbers. 
Conjugate fields have opposite $U(1)' \times U(1)^4$
quantum numbers.
The fields $\Delta_i$ and $T_i$ are tabulated in terms
of their $U(1)' \times SO(10) \times SO(6) \times U(1)^4$
quantum numbers. }
}
\end{table}
 
As can be seen, the matter and
Higgs fields in this string model carry additional charges under
additional U(1) symmetries~\cite{aehn}. There exist various
singlet fields, and  hidden-sector 
matter fields which transform
non-trivially under the SU(4) $\times$ SO(10) gauge symmetry,
some as sextets under SU(4), namely $\Delta_{1,2,3,4,5}$, and some as
decuplets under SO(10), namely $T_{1,2,3,4,5}$. There are also
quadruplets of the hidden SU(4) symmetry which possess fractional
charges. However,  these are confined and
do not concern us further.

The usual flavour assignments of the light
Standard Model particles in this model are as follows:
\bea
\bar{f}_1 : \bar{u}, \tau, \; \; \;
\bar{f}_2 : \bar{c}, e/ \mu, \; \; \;
\bar{f}_5 : \bar{t}, \mu / e , \nonumber \\
F_2 : Q_2, \bar{s}, \; \; \;
F_3 : Q_1, \bar{d}, \; \; \;
F_4 : Q_3, \bar{b} , \nonumber \\
\ell^c_1 : \bar{\tau}, \; \; \;
\ell^c_2 : \bar{e}, \; \; \;
\ell^c_5 : \bar{\mu} ,
\label{assignments}
\eea
up to mixing effects which are discussed in more detail in
\cite{ELLN}.
We chose non-zero vacuum expectation values for the 
following singlet and hidden-sector fields:
\beq
\Phi_{31}, \bar{\Phi}_{31}, \Phi_{23}, 
\bar\Phi_{23},\phi_2, \bar{\phi}_{3,4},  \phi^-, 
 \bar\phi^+ ,  \phi_{45},  \bar{\phi}_{45}, \Delta_{2,3,5}, T_{2,4,5} .
\label{nzv}
\eeq
The vacuum expectation values of the hidden-sector fields
must satisfy the additional constraints 
\beq
 T_{3,4,5}^2=T_i\cdot T_4=0,\,\, \Delta_{3,5}^2=0,\,\, T_2^2+\Delta_2^2=0 .
\label{hcon}
\eeq
For further discussion, see~\cite{ELLN} and references therein.

{\bf Acknowledgements.} 
M.E.G and S.L. are grateful to 
CERN for hospitality and support. 
S.L. would also like to thank the University of Huelva 
for kind hospitality. 
The research of S. Lola is co-funded by the FP6 Marie Curie Excellence
Grant MEXT-CT-2004-014297.  Additional support for research visits
has been provided by the European Research 
and Training Network MRTPN-CT-2006 035863-1
(UniverseNet) and by the Greek Ministry of Education EPAN program, B.545. 
M.E.G. acknowledges support from the `Consejer\'{\i}a de Educaci\'on de 
la Junta de Andaluc\'{\i}a', the Spanish DGICYT under contracts 
BFM2003-01266, FPA2006-13825 and the European Network for Theoretical 
Astroparticle 
Physics (ENTApP), member of ILIAS, EC contract number
RII-CT-2004-506222.


\begin{thebibliography}{99}

\bibitem{S-Kam}
Y. Fukuda et al., Super-Kamiokande Collaboration, 
Phys. Rev. Lett. 81 (1998) 1562, Phys. Rev. Lett. 82 (1999) 1810
and Phys. Rev. Lett. 82 (1999) 2430.

\bibitem{SNO}
Q. R. Ahmad et al., SNO Collaboration, Phys. Rev. Lett. 87 (2001) 071301
and  Phys. Rev. Lett. 89 (2002) 011301.

\bibitem{KamLAND}
K. Eguchi et al., KamLAND Collaboration, Phys. Rev. Lett. 90 (2003) 021802;
T. Araki et al., KamLAND Collaboration, Phys. Rev. Lett. 94 (2004) 081801.

\bibitem{K2K}
M.H. Ahn et al., K2K Collaboration, Phys. Rev. Lett. 90 (2003) 041801.

\bibitem{MINOS}
D.G. Michael et al., MINOS colaboration, Phys. Rev. Lett. 97 (2006) 191801. 

\bibitem{N-data}
For an extensive list of references on the neutrino oscillation, reactor and
accelerator data, and for related global fits,
see: M. Maltoni, T. Schwetz, M.A.Tortola, J.W.F. Valle,
New J. Phys. 6 (2004) 122; 
M.C. Gonzalez-Garcia, Phys. Scripta T121 (2005) 72.

\bibitem{MSW}  See, for example, L. Wolfenstein, Phys. Rev. D17 (1978)
20; S.P. Mikheyev and A.Yu. Smirnov, Yad. Fiz. 42 (1985) 1441 and
Sov. J. Nucl. Phys. 42 (1986) 913.

\bibitem{CHOOZ}
M. Apollonio et al., CHOOZ Collaboration, 
Phys. Lett. B338 (1998) 383; Phys. Lett. B420 (1998)
397.

\bibitem{leptorev}
For a review see W. Buchmuller, 
lectures at ESHEP 2001, Beatenberg, Switzerland,
{\it Preprint} hep-ph/0204288.

\bibitem{BM}
F. Borzumati and A. Masiero, Phys. Rev. Lett. 57 (1986) 961.

\bibitem{KO}
For  reviews, see: Y. Kuno and Y. Okada, 
Rev. Mod. Phys. 73 (2001) 151, J. Aysto et al.,hep-ph/0109217,
{\it Report of the Stopped Muons Working Group for the ECFa-CERN study
on Neutrino Factory and Muon Storage Rings at CERN.}


\bibitem{LEP-CP-NU}
G.C. Branco, R. Gonzalez Felipe, F.R. Joaquim and M.N. Rebelo,
Nucl.Phys. B640 (2002) 202;
G.C. Branco, R. Gonzalez Felipe, F.R. Joaquim, 
I. Masina, M.N. Rebelo and C.A. Savoy,
Phys. Rev. D67 (2003) 073025;
J. Ellis, J. Hisano, S. Lola and M. Raidal,
Nucl.Phys. B621 (2002) 208;
J. Ellis and M. Raidal, Nucl.Phys. B643 (2002) 229;
T. Endoh, S. Kaneko, S.K. Kang, T. Morozumi, M. Tanimoto,
Phys.Rev.Lett.89 (2002) 231601;
S. Pascoli, S.T. Petcov and
W. Rodejohann, Phys.Rev.D68 (2003) 093007;
L. Velasco-Sevilla, JHEP 0310 (2003) 035.

\bibitem{FN}  C. D. Froggatt and H. B. Nielsen, Nucl. Phys. B147 (1979)
277.

\bibitem{IR}  L. Ibanez and G.G. Ross, Phys. Lett. B332 (1994) 100.


\bibitem{guid}
A very large number of netrino mass and mixing patterns has been proposed
in the literature. For a review, see 
G. Altarelli and F. Feruglio,
Phys. Rept. 320 (1999) 295, and references therein.

\bibitem{LR}
See for instance
S. Lola and G.G. Ross, Nucl. Phys. B553 (1999) 81, and references therein.

\bibitem{nonab}
See for instance
Y. L. Wu, Phys. Rev. D59 (1999) 113008;
C. Wetterich,  Phys. Lett. B451 (1999) 397;
I. de Medeiros Varzielas, S.F. King and G.G. Ross,
Phys. Lett. B644 (2007) 153 and
hep-ph/0607045.

\bibitem{nonab-KR}.
S.F. King and G.G. Ross, 
Phys. Lett. B520 (2001) 243 and
Phys. Lett. B574 (2003) 239.

\bibitem{neutrino_rev}
For a review see R.~N.~Mohapatra {\it et al.}, hep-ph/0510213.

\bibitem{REAP}
S. Antusch, J. Kersten, M. Lindner, M. Ratz and M.A. Schmidt,
JHEP 0503 (2005) 024.

\bibitem{leptogenesis}
L. Covi, E. Roulet and F. Vissani,
Phys.  Lett. B384 (1996) 169;
M. Plumacher, Nucl. Phys. B530 (1998) 207;
M. Flanz and E.A. Paschos, Phys. Rev.
D58, 113009 (1998);
A. Pilaftsis, Int. J. Mod. Phys. A14 (1999) 1811;
J. Ellis, S. Lola and D.V. Nanopoulos,
Phys. Lett. B452 (1999) 87; 
W. Buchmuller and M. Plumacher,
Int. J. Mod. Phys. A15 (2000) 5047;
C.H. Albright and S.M. Barr, Phys. Rev. D70 (2004) 033013;
G.C. Branco, M.N.Rebelo and J.I. Silva-Marcos,
Phys. Lett. B633 (2006) 345.


 
\bibitem{cehno1}B. Campbell, J. Ellis, J.Hagelin, D.V. Nanopoulos and
K.A. Olive,
Phys. Lett. {B197} (1987) 355.

\bibitem{JD}
J. Ellis, D.V. Nanopoulos and K. Olive,
Phys. Lett. B300 (1993) 121.

\bibitem{ELN}
J.R. Ellis, S. Lola and D.V. Nanopoulos, 
Phys. Lett. B452 (1999) 87.

\bibitem{Spergel:2003cb}
D.N. Spergel {\it et al.}, WMAP Collaboration,
Astrophys. J. Suppl. 148 (2003) 175.

\bibitem{fl-lep1}
 R.~Barbieri, P.~Creminelli, A.~Strumia and N.~Tetradis,
  Nucl.\ Phys.\  B {\bf 575} (2000) 61; 
T.~Endoh, T.~Morozumi and Z.~h.~Xiong,
 Prog.\ Theor.\ Phys.\  {\bf 111} (2004) 123; 
A.~Pilaftsis and T.~E.~J.~Underwood,
  Phys.\ Rev.\  D {\bf 72} (2005) 113001.

\bibitem{fl-lep2}
A. Abada {\it et al.}, JCAP 0604 (2006) 004 and JHEP 0609 (2006) 010;
O. Vives, Phys. Rev. D 73 (2006) 073006;
A. de Simone and A. Riotto, JCAP 0702 (2007) 005;
S. Blanchet and P. Di Bari, JCAP 0703 (2007) 018.


\bibitem{HIM} 
J. Hisano, T. Moroi, K. Tobe, M. Yamaguchi and 
T. Yanagida, Phys. Lett. B357 (1995) 579;
J. Hisano, T. Moroi, K. Tobe and M. Yamaguchi,
Phys. Rev. D53 (1996) 2442;  
M.E. Gomez and H. Goldberg, 
Phys. Rev. D53 (1996) 5244;
J. Hisano and D. Nomura,
Phys. Rev. D59 (1999) 116005.

\bibitem{LAM} 
S. Lavignac, I. Masina and C.A. Savoy, 
Phys. Lett. B520 (2001) 269 and Nucl. Phys. B633 (2002) 139.

\bibitem{EHRS} 
J. R. Ellis, J. Hisano, M. Raidal and Y. Shimizu,
Phys. Lett. B528 (2002) 86.

\bibitem{LFVap}
P.H. Chankowski, J.R. Ellis, S. Pokorski, M. Raidal
and Krzysztof Turzynski, Nucl.Phys. B690 (2004) 279.



\bibitem{K-SU5}
G.L. Kane, S.F. King, I.N.R. Peddie and L. Velasco-Sevilla,
JHEP 0508 (2005) 083.

\bibitem{SU5-a}
G. Altarelli and F. Feruglio, Phys. Lett. B451 (1999) 388.

\bibitem{King:2002nf}
  S. F. King,  JHEP 0209 (2002) 011.

\bibitem{tanb}
T. Banks, Nucl. Phys. B303 (1988) 172;
L.J. Hall, R. Rattazzi and U. Sarid, Phys. Rev. D50 (1994) 7048; 
M. Carena, M. Olechowski, S. Pokorski, and C.E.M. Wagner, 
Nucl. Phys. B426 (1994) 269; E.G. Floratos, G.K. Leontaris 
and S. Lola, Phys. Lett. B365 (1996) 149;
M.E. Gomez, T. Ibrahim, P. Nath and S. Skadhauge, Phys. Rev. D70 (2004) 035014.



\bibitem{ELLN}
J. Ellis, G.K. Leontaris, S. Lola and 
D.V. Nanopoulos,  Phys. Lett. B425 (1998) 86.


\bibitem{ELLN2}
J. Ellis, G.K. Leontaris, S. Lola and 
D.V. Nanopoulos, Eur. Phys. J. C9 (1999) 389.

\bibitem{aehn}
 I. Antoniadis, J. Ellis, J. Hagelin and D.V. Nanopoulos,
 Phys. Lett.  B194 (1987) 231 and
Phys. Lett.  B231 (1989) 65.

\end{thebibliography}
\end{document}